\pdfoutput=1
\documentclass[PAPER, atlasdraft=false, UKenglish, cernpreprint, texmf, orcidlogo]{atlasdoc}
\usepackage[subcaption]{atlaspackage}
\usepackage{atlasbiblatex}

\usepackage[bsm=true, particle=true, hepprocess=true]{atlasphysics}

\usepackage{threeparttable}

\graphicspath{{logos/}{figures/}}

\usepackage{ANA-SUSY-2022-15-PAPER-defs}

\usepackage{ANA-SUSY-2022-15-PAPER-preamble}

\AtlasTitle{Search for displaced decays of long-lived particles in events with
missing transverse momentum in $\sqrt{s} = 13$ TeV $pp$ collisions with the ATLAS detector}

\AtlasAbstract{%
A search for long-lived particles in events with significant missing transverse
momentum and at least one displaced vertex is presented. This analysis is performed
using 137 $\text{fb}^{-1}$ of $pp$ collision data collected between 2016--2018 during Run 2
of the Large Hadron Collider by the ATLAS detector. Displaced vertices are identified using two different
secondary vertexing algorithms, including a novel "fuzzy" vertexing algorithm optimized
for identifying displaced decays of heavy quarks. Separate searches are performed using
each algorithm, and the expected Standard Model background is independently estimated
for each search using a data-driven procedure. No significant excess is observed over
the background in either case. The results are used to set 95\% confidence-level limits on
potential beyond-the-Standard Model physics that could produce this final state.
Results are interpreted in the context of four models: long-lived
gluinos that form $R$-hadrons before decaying, neutralinos decaying via
Higgs-mediated channels in the Bino-Wino coannihilation model, long-lived Higgsinos decaying to axinos,
and an exotic Higgs portal model predicting displaced decays of light pseudoscalars.
}

\AtlasRefCode{SUSY-2022-15}

\PreprintIdNumber{CERN-EP-2026-052}

\AtlasJournal{JHEP}
\AtlasJournalRef{JHEP 08 (2026) 016}
\AtlasDOI{10.1007/JHEP08(2026)016}

\AtlasCoverEgroupAnalysisTeam{atlas-susy-2022-15-analysis-team@cern.ch}

\hypersetup{pdftitle={ATLAS document},pdfauthor={The ATLAS Collaboration}}

\begin{document}

\maketitle

\tableofcontents

%
%

\section{Introduction}
\label{sec:intro}

As the search continues for new physical processes that go beyond what is described by the Standard Model (SM) of particle physics -- so-called \textit{BSM physics} -- the range of hypothesized BSM models and their observable signatures has also continued to widen. The classes of models that predict the presence of long-lived particles (LLPs) are particularly interesting both theoretically and experimentally. Theoretically, long lifetimes are generated through the presence of heavy, off-shell particles in the decay, through phase space suppression factors or suppression from approximate symmetries that introduce very small effective coupling constants~\cite{Knapen:2022afb}. Experimentally, such particles may have proper lifetimes ranging from a few picoseconds to several nanoseconds, resulting in decay lengths from fractions of a millimeter to several meters or more, thus leading to diverse signatures involving multiple types of sensors and detector apparatus.

LLP decays can yield a variety of characteristic experimental signatures depending on the production mechanisms involved, the particle kinematics, and the properties of the decay process. One of the most striking signatures at a collider experiment is the presence of a massive, identifiable displaced vertex (DV).
This remarkable signature can be produced by decays that involve multiple outgoing charged particles and occur within the fiducial volume of the detector, but at a significant distance from the interaction point (IP) of the incoming beams.
While such displaced signatures of BSM processes have long been sought after at both colliders and beam dump experiments, the energy, intensity, and data volume of the Large Hadron Collider (LHC)~\cite{Evans:2008zzb}, along with the sophistication and sensitivity of detectors like those of the ATLAS experiment~\cite{PERF-2007-01}, provide unprecedented sensitivity in searches for these rare yet distinctive processes.

The results presented here use data recorded by the ATLAS detector to conduct a broad search for evidence of BSM processes that yield DVs produced within the inner detector (ID) in conjunction with significant missing transverse momentum $\vec{p}_{\text{T}}^{\text{miss}}$, with magnitude \met, indicating the production of particles that subsequently escape the ATLAS detector unobserved.
Using the 2016-2018 (Run 2) dataset of \invfb{137} proton-proton (\pp) collisions collected by the ATLAS experiment at a center-of-mass energy of $\sqs = \tev{13}$~\cite{DAPR-2021-01}, this analysis searches for events with one or more DVs with a large reconstructed mass and a high track multiplicity.
The relatively inclusive set of selection criteria implements two different DV reconstruction algorithms, one of which is specifically optimized for cases in which LLP decays yield heavy-flavor hadrons. The resulting low-background search results are interpretable in the context of several BSM models that predict displaced signatures, including models never-before confronted with data such as a full supersymmetric axion/axino model.

This search expands the dataset used by over a factor of four compared to previous ATLAS searches for similar signatures~\cite{SUSY-2016-08}. In addition, ATLAS has searched for DVs in events with final states involving jets or muons~\cite{SUSY-2018-13, SUSY-2018-33, EXOT-2022-17}, and the CMS and LHCb Collaborations have also searched extensively for related signatures and models of BSM physics~\cite{CMS-SUS-14-020, CMS-EXO-16-003, CMS-EXO-17-018, CMS-EXO-18-007, CMS-EXO-19-013, CMS-EXO-19-021, CMS-EXO-20-009, CMS-EXO-22-020, LHCb:2017xxn, LHCb:2020akw, LHCb:2021dyu}.

The details of the physics models used for the interpretation of the results are described in \Sect{\ref{sec:signals}}.
\Sect{\ref{sec:atlas}} presents the relevant information regarding the data taking conditions and detector apparatus used in the analysis.
The special reconstruction algorithms and event selection criteria are presented in \Sect{\ref{sec:reco}}.
\Sect{\ref{sec:sr_backgrounds}} discusses the detailed event selection criteria and methodology used to estimate the sources and levels of backgrounds relevant to the search.
The sensitivity to experimental and theory uncertainties of the analysis is described in \Sect{\ref{sec:systematics}}.
Finally, \Sect{\ref{sec:results}} presents the results and their interpretations.


%
\section{Signal models and Monte Carlo simulation}
\label{sec:signals}

Supersymmetric (SUSY) extensions of the SM of particle physics provide solutions to the hierarchy problem~\cite{Sakai:1981gr,Dimopoulos:1981yj}
by predicting new particles which are partners of SM particles~\cite{Golfand:1971iw_updated,Volkov:1973ix,Wess:1974tw,Wess:1974jb,Ferrara:1974pu, Salam:1974ig} -- the partners of SM fermions being bosons and those of SM bosons being fermions. $R$-parity-conserving SUSY models~\cite{Farrar:1978xj} are considered in this paper. In these models, the supersymmetric partners of the Higgs and electroweak gauge bosons mix to form mass eigenstates known as neutralinos (\ninoi) and charginos (\chargoi), which are linear combinations of the Bino, Wino, and Higgsino gauge eigenstates. In most of the models considered here, the lightest neutralino (\ninoone) is the stable lightest supersymmetric particle (LSP) and a dark-matter candidate~\cite{Goldberg:1983nd,Ellis:1983ew}, though in the Dine-Fischler-Srednicki-Zhitnitsky (DFSZ) axino model the fermionic superpartner of the axion (the axino, \axino) serves as the LSP~\cite{Hoshino:2025lqz} (see \Sect{\ref{sec:axino}} for more details).

Four BSM LLP models are considered that involve a variety of production mechanisms and decay modes.
These consist of a \RhadronModel model~\cite{Giudice:2004tc}, long-lived neutralinos in a \BinoWinoModel model~\cite{Nagata_2015,Rolbiecki_2015}, Higgsinos decaying to axinos in the \AxinoModel model, and light \scalars from an exotic \HiggsPortalModel model~\cite{ARCADI20201}.
Despite their different theoretical motivations, these models share common experimental signatures: DVs and significant \MET. This commonality allows a single general search strategy to be used.
For optimized sensitivity, different signal regions are constructed for the different topologies across these models.
Moreover, these models cover a wide range of both parameter space and physics motivation, from the pair-production of long-lived gluinos via the strong interaction, to axion-axino physics, to an exotic Higgs portal model.

To maximize sensitivity to this range of models, two classes of signal regions (SRs) are designed: one optimized for processes that typically produce one or more well-separated massive DVs per event, and a second optimized for models with LLP decays to $b$-quarks. The details of these standard DV (SDV) and ``fuzzy'' DV (FDV) SRs, respectively, are given in \Sect{\ref{sec:sr_backgrounds}}.

The dominant background for these DV+\MET signatures is expected to be $Z(\rightarrow\nu\nu)$+jets processes in which the \MET due to the neutrinos occurs in conjunction with anomalously massive DVs from hadronic interactions with detector materials. Consequently, these backgrounds can only be estimated using precise information regarding the position and composition of detector materials as well as the particle interaction dynamics with those materials.
Modeling of these background processes faces significant challenges, such as the misreconstruction of tracks, in even the most modern Monte Carlo (MC) simulations. This leads to poor simulation-based modeling of the kinematics and number of DVs produced by SM processes, thus necessitating a data-driven approach for background estimation.
Nonetheless, both \Zjets and $\gamma$ + jets MC samples are used for preliminary studies of the analysis design and the selection
requirements.

\begin{figure}[!htb]
\begin{center}
\subfloat[\label{fig:feynman-RHadron}]{
\includegraphics[width=0.45\textwidth]{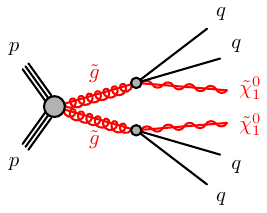}
}
\hfill
\subfloat[\label{fig:feynman-BinoWino}]{
\includegraphics[width=0.45\textwidth]{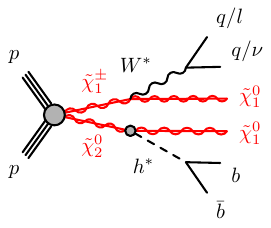}
}\\
\subfloat[\label{fig:feynman-Axino}]{
\includegraphics[width=0.45\textwidth]{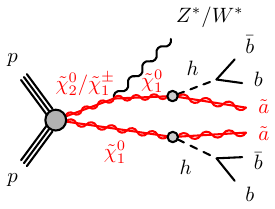}
}
\hfill
\subfloat[\label{fig:feynman-HiggsPortal}]{
\includegraphics[width=0.45\textwidth]{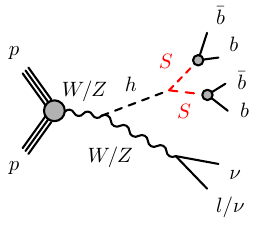}
}\\
\end{center}
\caption{Illustrative diagrams for the four signal interpretations considered in this analysis; the \protect\subref{fig:feynman-RHadron} \RhadronModel model described in \Sect{\ref{sec:rhadron}}; \protect\subref{fig:feynman-BinoWino} the \BinoWinoModel model described in \Sect{\ref{sec:bino_wino}}; \protect\subref{fig:feynman-Axino} the \AxinoModel model described in \Sect{\ref{sec:axino}}; and \protect\subref{fig:feynman-HiggsPortal} the \HiggsPortalModel described in \Sect{\ref{sec:higgs_portal}}. Red lines indicate BSM particles, and the small gray circles indicate displaced vertices.}
\label{fig:feynman}
\end{figure}
%

\subsection{\RhadronModel}
\label{sec:rhadron}

The first of the models considered is a split-SUSY model~\cite{Giudice:2004tc} with pair-produced long-lived gluinos which hadronize into
\rhadrons~before decaying (see  \Fig{\ref{fig:feynman-RHadron}}). Such signatures can be created in a number of similar models~\cite{Farrar:1978xj, ArkaniHamed:2004fb}.
Each \rhadron decays to two SM quarks and a stable neutralino LSP, \ninoone, via an $R$-parity conserving process mediated by heavy virtual squarks, thus yielding significant \met in the final state as the neutralinos escape the detector undetected.
The decay of the gluino is suppressed by the high mass of the virtual squarks which mediate the decay, causing the gluino to become long-lived, resulting in up to two massive displaced vertices.
The branching ratio of the \rhadron is assumed to be flavor democratic to the first two generations of quarks.
This model is targeted by the SDV SR as the decay products are light quarks.


\subsection{\BinoWinoModel}
\label{sec:bino_wino}

The second benchmark model is the pair-production of the lightest chargino, \chinoonepm, and the long-lived second-lightest neutralino, \ninotwo, in the \BinoWinoModel model~\cite{Nagata_2015,Rolbiecki_2015}  (see  \Fig{\ref{fig:feynman-BinoWino}}).
In this model, \chinoonepm and \ninotwo are Wino-like, and the neutralino LSP, \ninoone, is Bino-like.
The mass difference between the \ninotwo and the \ninoone is favored to be a few tens of GeV to match experimentally observed dark matter relic density constraints~\cite{Nagata_2015}.
While the \chinoonepm decays promptly to the \ninoone and a pair of leptons or quarks via an off-shell \Wboson-boson emission in most of the parameter space, the \ninotwo may be long-lived depending on the parameter space.

When the Higgsino mass is large enough, and the ratio of vacuum expectation values is relatively small, which is the parameter space considered in this analysis, \ninotwo is typically long-lived and decays via an off-shell Higgs boson, which subsequently decays mostly to a pair of $b$-hadrons.
A single massive FDV is expected, due to the $b$-hadrons, along with \met from the escaping neutralino LSP.


\subsection{\AxinoModel}
\label{sec:axino}

Another model considered for this search is a SUSY scenario in which the minimal supersymmetric standard model (MSSM)~\cite{Djouadi:1998di} is extended with the addition of the DFSZ~\cite{dfsz_1980,dfsz_1981} axion and associated Peccei-Quinn (PQ) symmetry~\cite{Hoshino:2025lqz} (see \Fig{\ref{fig:feynman-Axino}} for a representative diagram). In this model, the Higgs sector carries PQ charge and so the axino has a direct coupling with both Higgsinos and SUSY fermions. The axino, as a neutral fermion, additionally has a small perturbative mixing with the up- and down-type Higgsino states, which gives rise to a fifth neutralino mass eigenstate \ninofive.

This analysis considers an $R$-parity conserving scenario in which the axino (or rather, \ninofive)
is both the lightest neutralino and the LSP, and so escapes the detector as a stable, undetected particle.
The next two lightest supersymmetric particles are \ninoone and \ninotwo, which are considered to be pure-Higgsino states with a very small mass gap. Note that while it is common for \ninoone to be reserved for the lightest neutralino state, \axino is chosen for the new, mostly-axino neutralino to avoid confusion with the standard MSSM neutralinos. The axino's couplings to other particles depend on the axion decay constant, \fa. Because the axion is only feebly coupled to the SM, \ninoone can be long-lived and its lifetime depends on both \fa and \mneutralino~\cite{Hoshino:2025lqz}.
The axino mass, \maxino, strongly influences the magnitude of the observable \met in the final state.

Due to the neutralino mixing, \ninoone can decay to \ninofive by emitting either a Higgs or a $\Zboson$ boson. This analysis considers $hh$ and $ZZ$ channels; the diagram in \Fig{\ref{fig:feynman-Axino}} is for the $hh$ channel. Only the $h \rightarrow b\bar{b}$ (with initially expected better sensitivity in the FDV SR) and $Z \rightarrow q\bar{q}$ (with expected better sensitivity in the SDV SR) decays are considered due to the requirements on the presence of high track-multiplicity DVs. These modes are enforced in the MC simulation by setting their branching ratios to 1, and the resulting cross-sections are scaled by the corresponding SM branching ratios to reflect the expected physical rates.


\subsection{\HiggsPortalModel to long-lived particles}
\label{sec:higgs_portal}

The final model used to interpret the results of this search is a \HiggsPortalModel model~\cite{ARCADI20201} (see \Fig{\ref{fig:feynman-HiggsPortal}}).
This model introduces a new decay channel for the Higgs boson into a pair of \scalar LLPs, called $S$, which subsequently decay into \bbbar pairs.
One or two FDVs are thus expected.
$S$ is generated as a pseudoscalar although the analysis does not exploit its charge-parity properties.
If both \scalar LLPs decay within the ID, DVs can be reconstructed from their decay products.
The source of \met for this model depends on the production mechanisms for the Higgs boson.
One mechanism involves the decay of top quarks, produced in association with the Higgs boson, into $W$ bosons that subsequently decay leptonically.
This process is the dominant contribution for the majority of the lifetime range of interest due to its large \met from the kinematics of its escaping neutrinos.
The other involves vector bosons directly produced alongside the Higgs boson, which also decay leptonically.
If one of the LLPs escapes detection in the calorimeter or is not reconstructed, it can also result in \met{} and leave only a single DV in the event.

%


%

\subsection{Monte Carlo simulations and event generation}
\label{sec:mc-generation}

MC simulations are used to model the expected signal processes and to study the event selection strategy (see \Sect{\ref{sec:sr_backgrounds}}). The signal processes described in \Sectrange{\ref{sec:rhadron}}{\ref{sec:higgs_portal}} are generated using either \MADGRAPH~\cite{Alwall:2014hca} or \POWHEG~\cite{Nason:2004rx, Frixione:2007vw}, with the decays of the primary particles provided by \MADSPIN~\cite{Artoisenet:2012st, Frixione:2007zp}, or by \Pythia~\cite{Sjostrand:2014zea}. Finally, \Pythia is used to simulate the parton showering and hadronization for all of the signal processes, and the subsequent decays of bottom and charm hadrons are performed by \EVTGEN~\cite{Lange:2001uf}.
The LLP decays are characterized by kinematic and lifetime distributions determined by the underlying BSM models. Background processes, including \Zjets and $\gamma$ + jets, are generated with \Sherpa~\cite{Gleisberg:2008ta,Bothmann:2019yzt}. All generated events for both signal and background processes are passed through a detailed \GEANT~\cite{Agostinelli:2002hh} simulation, which models the interactions of particles with the ATLAS detector, including the propagation of \rhadrons~\cite{ATL-PHYS-PUB-2019-019} and realistic collision conditions with pileup representative of each run period~\cite{SIMU-2020-01}. The simulated events then undergo standard ATLAS digitization, reconstruction, and an additional tracking step to recover displaced tracks from the signal, with two secondary vertexing algorithms applied to identify both standard and fuzzy displaced vertices (see \Sect{\ref{sec:reco}}).

Detailed descriptions of the implementation of each model are provided in the following subsections, while Table~\ref{tab:mc_generation} summarizes the technical details of MC event generation.

\begin{table}[h]
\centering
\caption{Summary of MC generation details for signal and background samples. \label{tab:mc_generation}}
\renewcommand{\arraystretch}{1.3}
\begin{tabular}{>{\raggedright\arraybackslash}p{0.16\textwidth}
>{\raggedright\arraybackslash}p{0.18\textwidth}
>{\raggedright\arraybackslash}p{0.16\textwidth}
>{\raggedright\arraybackslash}p{0.15\textwidth}
>{\raggedright\arraybackslash}p{0.22\textwidth}}
\toprule
Model & Generator & Decay simulation & Cross-section order (QCD) & Mass / Lifetime ranges \\
\midrule
\RhadronModel &
\MADGRAPH 2.6.2 \newline (NNPDF2.3LO) &
\PYTHIA 8.24 \newline (A14 tune) &
NNLO+NNLL &
\mgluino: 400--3000 GeV \newline
\dMass: 10--2900 GeV \newline
$\tau$: 0.01--30 ns \\
\midrule
\BinoWinoModel &
\MADGRAPH 2.9.5 \newline (NNPDF23LO) &
\MADSPIN + \PYTHIA 8.212 \newline (A14 tune) &
NLO+NLL &
\mslneutralino: 100--1100 GeV \newline
\dMassWinoBino: 20--50 GeV \newline
$\tau$: 0.01--3 ns \\
\midrule
\AxinoModel &
\MADGRAPH 2.9.16 \newline (NNPDF23LO) &
\MADSPIN + \PYTHIA 8.307 \newline (A14 tune) &
NLO &
\maxino: 0--500 GeV \newline
\mneutralino: 200--1000 GeV \newline
\maxion: 56--5690~\si{\micro\eV} \newline
$\tau$: 0.00058--6.0 ns \\
\midrule
\HiggsPortalModel &
\POWHEG 2 \newline (NNPDF23LO) &
\PYTHIA 8.307/8.212 \newline (A14 tune) &
NLO &
$m_S$: 5--55 GeV \newline
$\tau$: 0.0033--3.3 ns \\
\midrule
$Z \rightarrow \nu\nu$ + jets &
\SHERPA 2.2.11 &
\SHERPA 2.2.11 &
NLO & -- \\
$Z \rightarrow \mu\mu$ + jets &
\SHERPA 2.2.11 &
\SHERPA 2.2.11 &
NLO & -- \\
$\gamma$ + jets &
\SHERPA 2.2.2 &
\SHERPA 2.2.2 &
NLO & -- \\
\bottomrule
\end{tabular}
\end{table}

\subsubsection*{\RhadronModel model}

The matrix element (ME) calculation for the gluino pair-production was performed with \MADGRAPH 2.6.2 at leading-order (LO) precision~\cite{Alwall:2014hca}.
\PYTHIA 8.24 was then used to simulate the decays of the \rhadrons, the parton showering, and the underlying event~\cite{Sjostrand:2014zea}.
The A14 set of tuned underlying event and shower parameters~\cite{ATL-PHYS-PUB-2014-021} was used with the \NNPDF[2.3 LO]~\cite{Ball:2012cx} parton distribution function set to generate the \rhadron samples.
The gluino pair-production cross-sections and their uncertainties have been calculated at approximate next-to-next-to-leading order (NNLO) with next-to-next-to-leading logarithmic (NNLL) accuracy in accordance with the PDF4LHC15 recommendations~\cite{Beenakker:2014sma, Butterworth:2015oua}.
The gluinos in the simulated samples have mass \mgluino between $400$--\gev{3000}, mean proper lifetime $\tau$ between $0.01$--\ns{30}, and neutralino mass \mneutralino set from a range of \dMass between $10$--\gev{2900}.

\subsubsection*{\BinoWinoModel model}

The ME for the pair-production of Wino-like \chinoonepm \ninotwo is calculated using \MGNLO 2.9.5 at next-to-leading-order (NLO) precision with next-to-leading-logarithmic (NLL) order accuracy, with NLO+NLL production cross-sections calculated by the LHC SUSY cross-section working group \cite{Fuks:2012qx, Fuks:2013vua}.
The decay of \chinoonepm and \ninotwo are simulated using \MADSPIN~\cite{Artoisenet:2012st}.
The events are then interfaced with \PYTHIA 8.212 to model parton showering, hadronization, and the underlying events with A14 tune and \NNPDF[2.3 LO]~\cite{Ball:2012cx} parton distribution function sets.
The branching ratios $\chinoonepm \rightarrow \ninoone W^{*}$ and $\ninotwo \rightarrow \ninoone h^{*}$ are set to 100\%.
For an off-shell Higgs boson with mass between $20$--\gev{50}, the branching ratio $h^{*}\rightarrow b\bar{b}$ is predicted to be approximately 90\%~\cite{deFlorian:2016spz}, and is set to 100\% in this analysis for simplicity. Note that approximately 100\% of the decays yield visible decay products in the form of charm and tau.
The simulated samples have \mslneutralino{} between $100$--\gev{1100}, $\tau$ between $0.01$--\ns{3}, and \dMassWinoBino between $20$--\gev{50}.

\subsubsection*{\AxinoModel model}

\MGNLO 2.9.16 is used to generate the ME for neutralino (\ninoone and \ninotwo) pair-production, with a small $\Delta m = \gev{2.5}$ between the masses of \ninoone and \ninotwo. This
causes a prompt $\ninotwo \ra \ninoone \, Z^*(\nu\bar{\nu})$ decay to a very off-shell $Z^*$, followed by the subsequent decay of both \ninoone to axinos. Due to the small mass difference and low-momentum
decay products, it is assumed that hadronic and leptonic decays of the $Z^*$ will result in similar selection efficiencies and not otherwise impact
the analysis. This process is also indistinguishable from chargino-neutralino pair production, where the charged Higgsino decays via $\chinopm \ra \ninoone, W^*$
as shown in the representative diagram in \Fig{\ref{fig:feynman-Axino}}. The production cross-section is reweighted to the
inclusive NLO Higgsino pair production cross-section \cite{Fuks:2012qx, Fuks:2013vua}, which also includes these processes.

The mean Higgsino lifetime is determined by the axion decay parameter \fa.
This parameter is related to the \textit{axion} and \textit{neutralino} masses, but not the axino mass (which remains a free parameter).
This is consistent as the mass difference between the axino and Higgsino is always at least 400 GeV, $\mneutralino - \maxino > 400$~GeV.
The primary neutralinos in the simulated samples have masses between $200$--\gev{1000}, whereas the axino mass ranges from massless up to \gev{500}.
For the QCD axion model, the axion mass, \maxion, is determined from \fa, for which two values are adopted in this analysis: $\fa = 5\times 10^9~\GeV$ and $5\times 10^{10}~\GeV$.
The corresponding axion masses are thus $\maxion = \muev{113}$ and $\muev{1138}$ respectively.
The cases of subsequent $\ninoone \ra \axino h$ and $\ninoone \ra \axino Z$ decays are considered, which are performed with \MADSPIN, with decay branching ratios $h \ra \bbbar$ and $\Zboson \ra \qqbar$ respectively set to 100\%.
Parton showering, hadronization and underlying event generation is subsequently handled by \PYTHIA 8.307  \cite{Bierlich:2022pfr}, using the A14 tune and \NNPDF[2.3 LO] parton distribution function set.

\subsubsection*{\HiggsPortalModel model}

MC samples were generated for Higgs boson production associated with vector bosons (\VH), top-quark pairs (\ttH), and via gluon--gluon fusion (\ggH) to support the Higgs portal search.
\Wboson and \Zboson bosons from \VH production and from top quark decays can produce neutrinos in the event, contributing to the \met{}.
Due to this, the \ttH mode dominates the sensitivity at most lifetimes, with the higher cross-section \ggH mode only relevant
at higher lifetimes when there is a higher chance of LLPs escaping detection.
The ME were generated at NLO for \VH and \ggH samples and at LO for \ttH sample using \POWHEG 2, and then the events were interfaced with \PYTHIA 8.212 for parton showering and hadronization.
VBF production was not used due to its very low sensitivity, as the low transverse boost makes it difficult to reconstruct an LLP
with high enough \met{}.

For the Higgs portal process, \PYTHIA 8.307 was used to decay the Higgs bosons from these samples into a pair of \scalar LLPs $S$, which subsequently decay to \bbbar with a 100\% branching fraction. In both cases the \NNPDF[2.3 LO] parton distribution function set and
A14 tune are used, and samples are scaled to production cross-sections from the LHC Higgs working group~\cite{deFlorian:2016spz}.
Samples were generated with $S$ masses between $5$--\gev{55} and $\tau$ between $0.0033$--\ns{3.3}. The \ggH mode was only generated
at $\tau = \ns{3.3}$, where it has non-negligible efficiency.



\section{ATLAS detector and data acquisition}
\label{sec:atlas}

%
%

%

%
%

%
\newcommand{\AtlasCoordFootnote}{%
ATLAS uses a right-handed coordinate system with its origin at the nominal interaction point (IP)
in the center of the detector and the \(z\)-axis along the beam pipe.
The \(x\)-axis points from the IP to the center of the LHC ring,
and the \(y\)-axis points upwards.
Polar coordinates \((r,\phi)\) are used in the transverse plane,
\(\phi\) being the azimuthal angle around the \(z\)-axis.
The pseudorapidity is defined in terms of the polar angle \(\theta\) as \(\eta = -\ln \tan(\theta/2)\) and is equal to the rapidity
$ y = \frac{1}{2} \ln \left( \frac{E + p_z}{E - p_z} \right) $ in the relativistic limit.
Angular distance is measured in units of \(\Delta R \equiv \sqrt{(\Delta y)^{2} + (\Delta\phi)^{2}}\).}

The ATLAS experiment~\cite{PERF-2007-01} at the LHC is a multipurpose particle detector
with a forward--backward symmetric cylindrical geometry and a near \(4\pi\) coverage in
solid angle.\footnote{\AtlasCoordFootnote}
It consists of the ID surrounded by a thin superconducting solenoid
providing a \qty{2}{\tesla} axial magnetic field, electromagnetic and hadronic calorimeters, and a muon spectrometer.

The ID covers the pseudorapidity range \(|\eta| < 2.5\).
The high-granularity silicon pixel detector covers the vertex region and typically provides four measurements per track,
the first hit generally being in the insertable B-layer (IBL).
It is followed by the SemiConductor Tracker (SCT), which usually provides eight measurements per track.
These silicon detectors are complemented by the transition radiation tracker (TRT),
which enables radially extended track reconstruction up to \(|\eta| = 2.0\).
The TRT also provides electron identification information
based on the fraction of hits (typically 30 in total) above a higher energy-deposit threshold corresponding to transition radiation.

Lead/liquid-argon (LAr) sampling calorimeters provide electromagnetic (EM) energy measurements
with high granularity within the region \(|\eta|< 3.2\).
A steel/scintillator-tile hadronic calorimeter covers the central pseudorapidity range (\(|\eta| < 1.7\)).
The endcap and forward regions are instrumented with LAr calorimeters
for EM and hadronic energy measurements up to \(|\eta| = 4.9\).
The muon spectrometer surrounds the calorimeters and is based on
three large superconducting air-core toroidal magnets with eight coils each.
The field integral of the toroids ranges between \num{2.0} and \qty{6.0}{\tesla\metre}
across most of the detector.
The muon spectrometer includes a system of precision tracking chambers up to \(|\eta| = 2.7\) and fast detectors for triggering up to \(|\eta| = 2.4\).
The luminosity is measured mainly by the LUCID--2~\cite{LUCID2} detector, which is located close to the beampipe.

Events of interest are selected by the two-level ATLAS trigger system~\cite{TRIG-2016-01}. The first-level trigger is implemented in hardware and uses a subset of the detector information to accept events at a rate below \qty{100}{\kHz}. This is followed by a software-based trigger that reduces the accepted event rate to \qty{1.25}{\kHz} on average depending on the data-taking conditions. A software suite~\cite{SOFT-2022-02} is used in data simulation, in the reconstruction and analysis of real and simulated data, in detector operations, and in the trigger and data acquisition
systems of the experiment.


%
%

The events used in this analysis are from \pp collisions at $\sqs=\tev{13}$, recorded under stable beam conditions at the LHC during 2016--2018.
The dataset corresponds to an integrated luminosity of \invfb{137} with an uncertainty of $0.84\%$, requiring all detector systems to be recording good quality data~\cite{DAPR-2018-01}. Data from 2015 are not included due to the need for special large-radius track reconstruction (see \Sect{\ref{sec:lrt}}) which was only available from 2016.

The primary dataset was collected by triggers targeting large \met~\cite{TRIG-2019-04}.
During data taking, events were accepted when the trigger-level calorimeter-based \met exceeded thresholds ranging from $50-70~\gevonly$, with the specific value adjusted according to the instantaneous luminosity in each data-taking period.
The efficiency of these triggers reaches approximately 100\% for events in which the high-precision \met calculated in the offline reconstruction exceeds 180~GeV.
Because muons deposit minimal energy in the calorimeters, samples of $Z \rightarrow \mu\mu$ events provide a mechanism to emulate the \met signature and to validate the MC simulation modeling of the \met trigger performance.
A comparison of trigger efficiencies between data and simulation in $Z \rightarrow \mu\mu$ events shows agreement at the level of 1\% both in the portion of the turn-on region that is used ($150 < \met < 180$ GeV) and on the efficiency plateau ($\met > 180$ GeV).
This procedure builds confidence that the \met trigger behavior is well-modeled in the simulation and hence reliable for searches for BSM signals.
Additional validation was performed using signal MC samples, where the trigger efficiency was studied as a function of the high-precision offline \met, confirming accurate modeling for events with genuine missing transverse momentum.
Further details regarding event selections based on the \met are discussed in \Sect{\ref{sec:objects}}.

Auxiliary data samples used to validate the background estimation were selected using triggers requiring one photon~\cite{TRIG-2018-05}, with transverse momentum $\ptPhoton > \gev{140}$, or triggers requiring multiple jets~\cite{TRIG-2018-08}.



\section{Event reconstruction and displaced vertexing}
\label{sec:reco}

The analysis signature is one or more DVs in the presence of sufficiently large \met.
Two different algorithms are used in this analysis in two separate signal regions for the reconstruction of DVs from tracks: standard secondary vertexing (outlined in \Sect{\ref{sec:vertexing}}) used to target general scenarios such as the \RhadronModel scenario, and fuzzy secondary vertexing (outlined in \Sect{\ref{sec:fuzzy}}) used to target scenarios with LLPs eventually decaying into $b$-hadron products such as in the \BinoWinoModel or \HiggsPortalModel scenarios.
The material map veto, used to select against vertices that are likely created as the result of hadronic interactions of particles with detector material, is an important aspect of the control, validation, and signal region definitions of the analysis.
Its design is outlined in \Sect{\ref{sec:material_map}}.
Definitions of other physics objects such as the \met used for the trigger defining the signal and validation regions, photons used for the trigger defining the control regions, as well as jets used for background estimation, are detailed in \Sect{\ref{sec:objects}}.
The specific cleaning requirements for tracks and vertices, along with the estimation of their background processes, are detailed along with analysis selections in \Sect{\ref{sec:sr_backgrounds}}.

%
%

%
\subsection{Tracks and primary vertices}
\label{sec:lrt}

ATLAS uses two tracking algorithms to reconstruct the trajectories of charged particles in the ID.
The first is the standard track reconstruction algorithm, which is optimized to efficiently reconstruct tracks originating either from the IP or from the decays of SM particles such as $b$-hadrons~\cite{PERF-2015-08}.
The standard tracking algorithm is performed in two passes, both of which place constraints on the track hit multiplicities and the transverse impact parameter (\d0) and longitudinal impact parameter (\z0) of the track candidates with respect to the primary vertex ($|\d0| < \qty{10}{\mm}$ and $|\z0| < \qty{250}{\mm}$, respectively).
The first pass forms silicon-seeded track candidates from hits in the pixel detector and SCT.
Hits in the TRT that are compatible with an extrapolated track candidate are then added to the track.
The second pass utilizes unused track segments in the TRT to define nearby seed regions, in which additional track seeds based on silicon hits are produced;
from these, silicon-seeded tracks are produced and then extended into the TRT, much as in the first pass.
The second tracking pass is used to recover reconstruction efficiency for tracks that have a slightly displaced origin and therefore a relatively small transverse impact parameter, such as tracks from $b$-decays.

The selections required by the standard track reconstruction algorithm result in inefficiencies for tracks with large impact parameters, which are critical for reconstructing the decays of LLPs.
In order to recover efficiency for tracks originating from LLP decays, a second track reconstruction algorithm called \textit{large-radius tracking} (LRT) is performed in a dedicated reconstruction step~\cite{ATL-PHYS-PUB-2017-014}.
Hits in the ID that are not associated to a standard track are used by the LRT algorithm to reconstruct track candidates with looser requirements on the transverse ($|\d0| < \mm{300}$) and longitudinal ($|\z0| < \mm{1500}$) impact parameters.
Unlike in standard tracking, LRT track candidates do not require hits in the pixel detector.
No distinction between tracks reconstructed by the standard and those reconstructed by the LRT algorithms is made in the subsequent reconstruction steps.

Collision vertices (CVs) consistent with $pp$ interactions from the beam are reconstructed using the tracks identified by the standard tracking algorithms~\cite{ATL-PHYS-PUB-2015-026}.
CVs are required to have at least two associated tracks.
The vertex with the greatest sum of squared transverse momenta of the tracks associated with the vertex \sumptVertex, is identified as the \textit{primary vertex} (PV).
Tracks are then refitted using this identified PV as a constraint to obtain the track parameters with respect to the point of closest approach in the transverse plane between a track and the beam line, or the perigee.

\subsection{Standard secondary vertices}
\label{sec:vertexing}

LLP decays in the ID resulting in two or more tracks can be reconstructed as a DV by a dedicated standard displaced vertexing algorithm~\cite{ATL-PHYS-PUB-2019-013}, and the resulting DVs are denoted as SDVs.
Displaced vertexing begins by forming two-track seed vertices from pairs of \textit{selected tracks}.
Selected tracks, including both standard and LRT tracks, must have $\pt > \gev{1}$ and pass several hit multiplicity requirements.
To be used in SDV seeding, tracks must have at least six SCT hits or at least one pixel hit, and tracks with fewer than two pixel hits must have at least one hit in the TRT.
Tracks with $\pt < \gev{25}$ are subject to additional requirements in order to suppress contributions from \textit{fake tracks} -- those erroneously produced via combinatorial reconstruction effects -- which tend to have low \pt.
These tracks are required to have at least seven SCT hits, and they must additionally have at least 20 hits in the TRT if they are reconstructed with $|\eta| < 1.7$.

Two-track seed vertices are formed by a $\chi^2$-based fitting algorithm \cite{Kostyukhin:685551}.
Once all two-track seed vertices of selected tracks have been formed, the seed vertices are iteratively merged to form $n$-track vertices.
If a track is compatible with more than one vertex at this stage, it is associated to the vertex with which it has a more compatible fit.
If the distance between the estimated positions of a pair of vertices is not significant enough, determined using a significance estimator based on the position uncertainty of the vertices, the pair of vertices can be merged.
At this point, SDVs are composed strictly of selected tracks.
In the final step of the algorithm, additional tracks are attached to SDVs to recover efficiency lost due to the strict seed track requirements.
These \textit{attached tracks} have loosened hit multiplicity requirements relative to selected tracks.
Attachment is accepted only if the reduced $\chi^2$ is less than 20.

At this point, the final vertex fit using all the selected and attached tracks is performed and track parameters with respect to the associated SDV are re-calculated.
The number of tracks associated with a SDV is denoted as \nTracks, and the visible invariant mass of the SDV, denoted \msdv, is calculated from the four-momenta of its constituent tracks, under the assumption that each track was produced by a charged pion.


%
\subsection{Fuzzy secondary vertices}
\label{sec:fuzzy}

Since the SDV algorithm relies heavily on vertex fitting, which assumes that all particles from the LLP decay originate from a single point, a significant number of tracks fail to be reconstructed as a single vertex when the LLP decay involves $b$-hadrons, due to their sizable decay length. \textit{Fuzzy displaced vertexing} is a DV reconstruction algorithm targeting such LLP decays.
This new vertexing algorithm was developed specifically for this analysis and is being deployed for the first time in the ATLAS experiment.

The fuzzy displaced vertexing algorithm begins by forming two-track seed vertices from selected track pairs.
Prior to seed vertex formation, two levels of selection are applied: track-level preselection and track-pair preselection.
For track-level preselection, tracks must meet the following criteria: $\pt{} > \gev{1}$, at least two hits in the silicon detector, and no association with any CVs.
All combinations of preselected tracks are then evaluated for their likelihood of originating from an LLP using a multivariable analysis.
A Boosted Decision Tree (BDT) algorithm in TMVA package \cite{TMVA} is used to determine the scores indicating the signal-track-pair-likeness using track pairs from an LLP in the signal MC events as the signal training samples and pairs of two tracks randomly selected from the background events as the background training samples.
The LLP used is \ninotwo{} in the \BinoWinoModel, with a mass of 300--500~GeV, a lifetime of 0.01--1~ns, and a mass difference from \ninoone{} of 30~GeV.
The input variables used for signal-track-pair-likeness BDT are listed in \Tab{\ref{tab:BTDVariables}}.
The importance of each input variable in determining the signal-track-pair-likeness slightly depends on the LLP lifetime, but generally, the most important input variables are the angular separations between two tracks in the $\eta$ and $\phi$ directions, $\Delta \eta$ and $\Delta \phi$.
The next important input variables are the transverse and longitudinal impact parameters of track \#1/\#2, $\d0^{(1/2)}, \z0^{(1/2)}$, the longitudinal impact parameter with respect to the beam spot of track \#1/\#2, $\z0^{\mathrm{BS}, 1/2}$, and the $\pt{}$ asymmetry between tracks, $\pt^\mathrm{asym}$.
This track-pair selection step ensures that seed vertex fitting is performed only on pairs likely to originate from an LLP, reducing computational demands and minimizing the formation of fake vertices.
A two-track seed vertex is created by the fitting algorithm used in the SDV algorithm if the score quantifying the signal-track-pair-likeness exceeds a defined threshold.
This threshold corresponds to approximately 98\% efficiency in selecting signal track pairs for the \BinoWinoModel model with $\mslneutralino = \gev{300}, \dMassWinoBino = \gev{30}$, and $\tau = \ns{0.3}$.
\begin{table}
\centering
\caption{Input variables to the signal-track-pair-likeness BDT. The superscript indicates the index of the track (1 or 2). Since the first seven rows include the variables for both tracks \#1 and \#2, there are 18 variables in total.}
\label{tab:BTDVariables}
\begin{tabular}{ll} \hline \hline
Variables & Description \\ \hline
$\eta^{(1)}, \eta^{(2)}$ & $\eta$ direction of tracks \#1 and \#2 \\
$\d0^{(1)}, \d0^{(2)}$ & Transverse impact parameter of tracks \#1 and \#2 \\
$S(\d0^{(1)}), S(\d0^{(2)})$ & $\d0$-significance of tracks \#1 and \#2 \\
$\z0^{(1)}, \z0^{(2)}$ & Longitudinal impact parameter of tracks \#1 and \#2 \\
$\z0^{\mathrm{BS}, 1}, \z0^{\mathrm{BS}, 2}$ & Longitudinal impact parameter with respect to the beam spot of tracks \#1 and \#2 \\
$S(\z0^{(1)}), S(\z0^{(2)})$ & $\z0$-significance of tracks \#1 and \#2 \\
$\mathcal{HC}^{(1)}$, $\mathcal{HC}^{(2)}$ & Category indicating the innermost hit layer of tracks \#1 and \#2 \\
& (0: First Pixel layer, 1: Second Pixel layer, 2: Third or fourth Pixel layers, 3: SCT layers) \\
$\Delta \eta$ & Angular separation in $\eta$ direction between tracks \#1 and \#2 \\
$\Delta \phi$ & Angular separation in $\phi$ direction between tracks \#1 and \#2 \\
$\pt^\mathrm{asym}$ & \pt asymmetry between tracks \#1 and \#2 (= $(\pt^{(1)} - \pt^{(2)})/(\pt^{(1)} + \pt^{(2)})$) \\
$\Delta \mathcal{HC}$ & Difference of $\mathcal{HC}$ between tracks \#1 and \#2 \\ \hline \hline
\end{tabular}
\end{table}

After forming two-track seed vertices, multiple seed vertices may correspond to a single LLP decay. To address this, vertex merging is performed to combine these seed vertices into a single vertex.
When the LLP decay involves $b$-hadrons, the seed vertices tend to be spatially distributed, so this merging step is performed independently of the fitting process. Each seed vertex is assigned a score indicating its signal-track-pair-likeness, which is initially determined during the two-track seed-forming step. The merging process is carried out in descending order of this score.

Seed vertices with particularly high scores, with a threshold of 0.15, are designated as \textit{primary seeds}. Approximately 85\% of the reconstructed seed vertices meet the primary seed criteria for the \BinoWinoModel model with $\tau = \ns{0.3}$.
The seed vertices are referred to as \textit{substitute seeds} of a primary seed if the differences in their positions $x$, $y$, and $z$ from the primary seed (\dxSeed, \dySeed, and \dzSeed) are all within a threshold.
The threshold is defined as \dxSeed, \dySeed, \dzSeed < \mm{1} for $\seedR{} < \mm{23.5}$, \dxSeed, \dySeed, \dzSeed < \mm{3} for $\mm{23.5} \leq \seedR{} < \mm{33.5}$, and \dxSeed, \dySeed, \dzSeed < \mm{5} for $\seedR{} \geq \mm{33.5}$, where \seedR{} is the radial position of the primary seed.
If the primary and substitute seeds satisfy the criterion in \Eqn{\ref{eq:FVMerging}}, the substitute seed is merged with the primary seed:
\begin{equation}
\sum \pt{}^{\mathrm{prim}} + \sum \pt{}^{\mathrm{sub}} > \gev{3.0} \times (\phi_\mathrm{max} - \pi/2) + \gev{4.0}, \label{eq:FVMerging}
\end{equation}
where $\sum \pt{}^{\mathrm{prim}}$ and $\sum \pt{}^{\mathrm{sub}}$ represent the sum of the \pt{} of the two tracks included in the primary and substitute seeds, respectively, and $\phi_\mathrm{max}$ is the maximum angle between the tracks from the primary and substitute seeds. This merging procedure is performed in descending order of the primary seeds' scores.
This criterion is designed to prevent the merging of large numbers of fake seed vertices formed from unrelated random track pairs.
As a final step, seed groups are iteratively merged until no two groups share any common seeds. The resulting seed groups are then defined as FDVs.
For FDVs, the vertex position is defined as the average of the position of the two-track seed vertices associated with the FDV, and track parameters with respect to this point are obtained.
The uncertainty of the vertex position is also defined as the standard deviation of the associated two-track seed vertices.
The number of tracks associated with an FDV and the visible invariant mass of an FDV, denoted as \nFDVTracks and \mfdv, are defined in the same way as the SDV.
The number of two-track seed vertices associated with the FDV is denoted as \nFDVSeeds.

The primary advantage of FDVs is their ability to include more tracks compared to SDVs, particularly when the signal LLP decay involves $b$-hadrons. This results in significantly higher signal selection efficiency.
The overall fake rate of FDVs is roughly comparable with that of SDVs but has different dependencies on the number of pileup collisions and the radial displacement from the beam axis $\rdv = \sqrt{x_\mathrm{DV}^2 + y_\mathrm{DV}^2}$.
The fake rate of FDVs with $\nFDVTracks \geq 3$ is approximately twice that of SDVs in the region with 33 mm < \rdv{} < 122 mm, roughly corresponding to the inside of the Pixel Detector, and about 1/5 in the region with \rdv{} > 122 mm, the outside Pixel Detector.
In the region with \rdv{} < 33 mm, inside the innermost Pixel layer, it can vary from a couple of times to about 20 times, depending on the number of pileup interactions.

\begin{figure}[!htb]
\centering
\subfloat[\label{fig:vertex_eff_rhadron}]{
\includegraphics[width=0.49\textwidth]{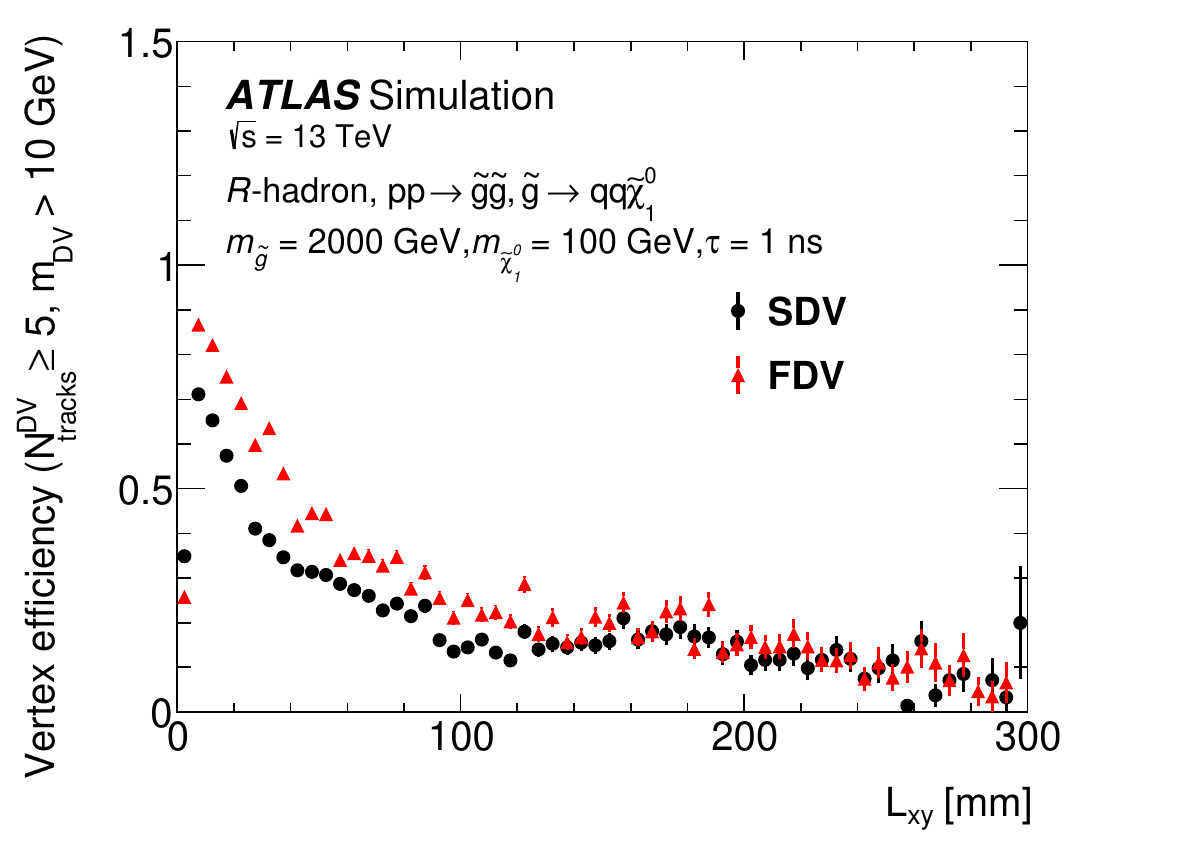}
}
\subfloat[\label{fig:vertex_eff_binowino}]{
\includegraphics[width=0.49\textwidth]{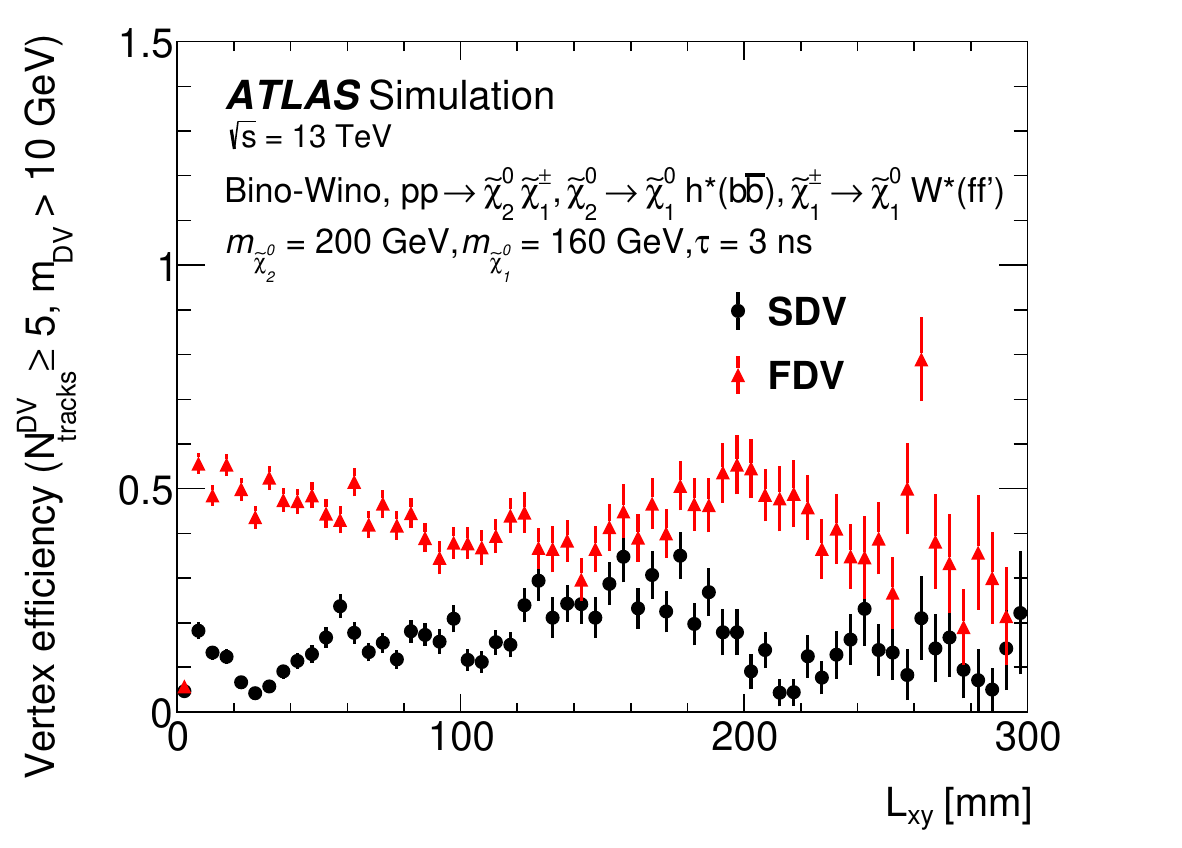}
}
\caption{Vertex reconstruction efficiency for the standard and fuzzy vertexing methods for \protect\subref{fig:vertex_eff_rhadron} \RhadronModel and \protect\subref{fig:vertex_eff_binowino} \BinoWinoModel models. $L_{xy}$ is the radial displacement of the truth position of the DV from the beamline axis. The spike around $L_{xy}$ = 260~mm is due to statistical fluctuations, as confirmed by looking at other signal models.
}
\label{fig:vertex_eff}
\end{figure}

\Fig{\ref{fig:vertex_eff}} illustrates the efficiency of reconstructing an LLP decay as a vertex with at least five tracks and an invariant mass greater than 10 GeV, using both standard and fuzzy displaced vertexing methods as a function of the truth radial displacement of the LLP decay from the beamline axis ($L_{xy}$).
Only tracks matched to the truth particles originating from the LLP decay are counted.
The fuzzy vertexing method demonstrates significantly higher efficiency than the standard method for the \BinoWinoModel model, e.g., 2.6 times at $L_{xy}$ = 50~mm and 3.2 times at $L_{xy}$ = 100~mm, whereas the increase in efficiency in the \RhadronModel model is limited.

%
\subsection{Detector material map}
\label{sec:material_map}

Hadronic interactions with detector material are a significant source of background in DV searches. To mitigate this, a veto is employed to exclude vertices reconstructed at positions consistent with being inside detector material, as identified by a material map~\cite{SUSY-2018-33}. DVs must pass this veto to be included in the signal regions, ensuring the search remains focused on genuine LLP decay signals rather than material-related background.

In addition to its role in defining the signal regions, the material map veto is also used to create a background-dominated validation region (VR) for background estimates by reversing the veto. DVs identified as originating inside detector material are dominated by background from hadronic interactions and have negligible signal contamination. This region provides a crucial sample for validating background estimate performance, as detailed in \Sects{\ref{sec:1FDVValidation}}{\ref{sec:2FDVValidation}}.

\begin{figure}[!htb]
\centering
\includegraphics[width=0.49\textwidth]{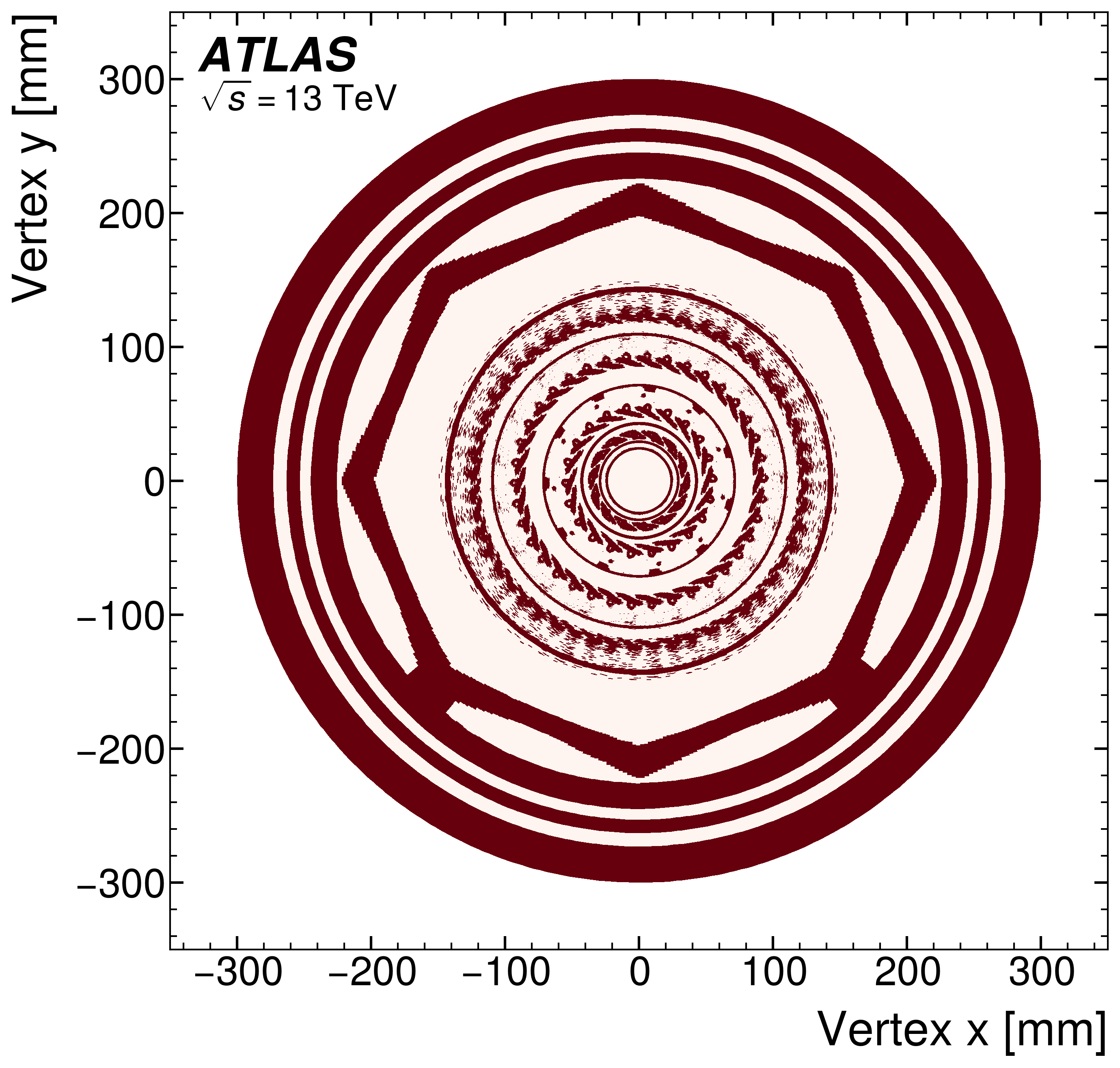}
\caption{Map showing a cross-section at $z=0$ of the physical material comprising the ATLAS detector's beam pipe
and tracking detector, used to define the "material map veto" \cite{SUSY-2018-33}.}
\label{fig:material_veto}
\end{figure}

The material map itself is constructed by analyzing the positions of low-mass vertices in data to identify regions of high vertex density. These regions are compared with the known locations of detector components to map the material distribution. The material map is represented as a cylindrical histogram, where each bin (sub-mm) is classified as either inside or outside material, as shown in \Fig{\ref{fig:material_veto}}. This map defines slightly less than 50\% of the ID's fiducial volume (radius $< 300$~mm) as subject to the veto criteria.

Two versions of the material map veto are implemented to provide flexibility. The loose veto checks whether the position of a DV falls within a bin identified as containing material and vetoes it accordingly. The strict veto excludes DVs not only based on their position but also by considering neighboring bins to account for potential mismatches between bin boundaries and actual material boundaries. Additionally, the strict veto incorporates reconstruction uncertainties in the DV position by excluding DVs whose uncertainty ellipsoid intersects any material-containing bins. This ensures a more robust rejection of material-related backgrounds for the FDV case as the vertex position is less precisely defined. A cross-check is also performed against the simulation geometry to validate the veto.
The strict veto is used for any SR requirements, whereas the loose veto is used in some CRs or VRs when required to reduce signal contamination.
In MC simulations, the same material map and veto algorithm are applied. However, in real data, the layers of the pixel detector are slightly offset in the $x-y$ plane due to their attachment to the beam pipe, the position of which can vary relative to the rest of the ATLAS detector. Such relative shifts are corrected for in the simulation to align with the real detector geometry.

By combining the material map veto for signal regions with the reversed veto for background validation, this approach ensures both an effective background suppression for LLP searches and reliable validation of background estimates.

%
%
\subsection{Physics object definitions}
\label{sec:objects}

Tracks and PVs are reconstructed as described in Section \ref{sec:lrt}.
Events are required to have the PV associated with at least two standard tracks with transverse momentum $\pT > \SI{500}{\MeV}$.
Tracks reconstructed by the standard track reconstruction algorithm (standard tracks) must have transverse momentum $\pT > \SI{500}{\MeV}$ and pseudorapidity range $|\eta|< 2.7$.
Tracks reconstructed by the LRT (LRT tracks) must have transverse momentum $\pT > \SI{1}{\GeV}$ and pseudorapidity range $|\eta|<5.0$ \cite{ATL-PHYS-PUB-2017-014}.

Muon candidates are reconstructed within the pseudorapidity range $|\eta|<2.5$ from tracks in the muon spectrometer that are matched to ID tracks. These candidates must have transverse momentum $\pT > \SI{10}{\GeV}$ and satisfy the \textit{medium} identification criteria detailed in Ref.~\cite{MUON-2018-03}.

Electron candidates are reconstructed from isolated energy deposits in the EM calorimeter matched to ID tracks~\cite{EGAM-2018-01,EGAM-2021-01}. They are required to satisfy $|\eta|<2.47$, $\pT > \SI{10}{\GeV}$, and the \textit{LooseAndBLayer} identification requirement described in Ref.~\cite{PERF-2017-01}.

Photons are used in the baseline vertex selections to remove displaced vertices arising from photon conversion, and are also used to define the control region.
Photon candidates are reconstructed from energy deposits in the EM calorimeter.
Both unconverted photon candidates without matching ID tracks and converted photon candidates with a two-track conversion vertex or with a single track are used.
They are required to satisfy $|\eta|<2.37$, $\pT > \SI{25}{\GeV}$, and the \textit{Tight} identification requirement described in Ref.~\cite{EGAM-2018-01}.

Jets are reconstructed from three-dimensional topological energy clusters in the calorimeters~\cite{PERF-2014-07} using the \antikt clustering algorithm~\cite{Cacciari:2008gp, Fastjet} with radius parameter $R=0.4$. Only jets with $\pT > \SI{20}{\GeV}$ and $|\eta|<2.8$ are retained. Events are rejected if they contain any jet with $\pT > \SI{20}{\GeV}$ that fails basic quality requirements designed to suppress detector noise and non-collision backgrounds~\cite{ATLAS-CONF-2015-029}. To avoid double-counting of energy depositions in the calorimeter, jets within a distance $\Delta R = 0.2$ of an electron candidate are discarded.
Jets containing $b$-hadrons ($b$-jets) are used for background estimation.
They are identified using the \textit{DL1} $b$-tagging algorithm~\cite{FTAG-2018-01}.
It is a multivariate classification algorithm that combines information from the impact parameters of the tracks and the secondary and tertiary decay vertices in the jet.
The $b$-tagging working point with a 77\% efficiency is used in this analysis.

The \MET{} is defined as the magnitude of the negative vector sum of the transverse momenta of all identified physics objects (electrons, photons, muons, jets) and an additional soft term accounting for energy not associated with hard objects.
For an overview of how \met is computed, see \Refn{\cite{JETM-2020-03}}.
In order to reduce the effects of jets from pileup collisions on \MET{} resolution, a jet vertex tagger \cite{ATLAS-CONF-2014-018} and a forward jet vertex tagger \cite{ATL-PHYS-PUB-2015-034} are applied to the jets with $|\eta| < 2.4$ and $\pt{} < 60$ GeV and those with $2.5 < |\eta| < 4.5$ and $\pt{} < 60$ GeV, respectively, in the \MET{} calculation. These jet vertex tagger requirements are not applied to the jets used in the analysis.
The soft term is constructed from all tracks that are not associated with any physics object and are linked to the PV. This approach ensures that the \MET{} reflects the best available calibration of the high-\pt{} objects while preserving pileup independence in the soft term. Unless otherwise noted, \met selections are based on this definition in the offline analysis. An additional related quantity, referred to as hadronic \MET, is calculated similarly as the magnitude of the negative vector sum of the transverse momenta of all calorimeter energy clusters, which are assumed to originate from hadronic objects~\cite{JETM-2020-03}. This effectively reduces contributions from identified leptons and photons.



%
\section{Signal regions and background estimation}
\label{sec:sr_backgrounds}

SRs are defined by a set of selection criteria intended to enrich the fraction of signal events and minimize the fraction of background events retained. These criteria are optimized for sensitivity to the signal models introduced in \Sect{\ref{sec:signals}}, while limiting contamination from background processes.
To identify these DVs, the two different vertexing algorithms introduced in \Sect{\ref{sec:reco}} are used.
This search is implemented as a counting experiment in two distinct SRs defined by the vertexing algorithm used.
Events containing vertices reconstructed with the SDV algorithm are categorized in the SDV SR, while events reconstructed with the FDV algorithm are placed in the FDV SR.
The expected number of background events in each SR and VR is estimated using signal depleted photon-triggered control regions (CRs), which have the same SR or VR selection criteria, except the events are photon-triggered instead of \met triggered and the \met selection criteria are dropped.
This is because the mechanisms behind the origin of DVs are linked to jets and overall event track activity, and are therefore independent to whether or not a photon or \met is present in the event.
The estimate is validated in VRs, where the events must be \met triggered, however some other SR selection criteria related to vertex requirements are inverted to ensure a signal depleted but kinematically similar region to the SR.
This section describes the details of the SR, VR, and CR definitions, as well the background estimation methods.
The uncertainties involved in the background estimation, as well as other systematic uncertainties, are detailed in \Sect{\ref{sec:systematics}}.
An independent validation for the SDV SR is also done via a component-based background estimate which uses different CRs and VRs.
The SDV and FDV regions are designed to be sensitive to different signal topologies, so orthogonality is not enforced between the two vertexing regions.
However, while the SDV SR searches for at least one DV, the FDV SR is divided into two orthogonal SRs that require exactly one FDV and at least two FDVs.
A common set of event-level selections determines the events considered in both SRs.

%
%
%

%
\subsection{Common event-level selections}
\label{sec:eventlevel}
Data recorded by the ATLAS Detector were first filtered during reconstruction to facilitate the specialized processing necessary for LLP searches, including LRT, standard and fuzzy secondary vertexings as described in \Sect{\ref{sec:reco}}.
This filter included events passing the \met{} trigger, the multijet trigger, the single-lepton trigger, or the single-photon trigger.
For the \met{}-triggered events, an additional requirement of  $\met > \gev{180}$ on the topological cluster-based hadronic was imposed.
Events in the SRs are required to pass both the \met{} triggers and the hadronic \met{} condition included in this filter, and also have an offline reconstructed $\met > \gev{150}$.
Events are vetoed if affected by detector issues such as calorimeter noise or data corruption.
To ensure the event originates from a $pp$ collision in the detector's IP, a PV with $|z_{\textrm{PV}}| < \qty{200}{\mm}$ is required.
Non-collision background (NCB) contamination, such as beam halo \cite{Drozhdin:1370157,Drozhdin:304958,Bruce:1646958}, is suppressed by requiring that the leading jet in the event deposits no more than 80\% of its energy in a single calorimeter layer and no more than 96\% of its energy in the EM calorimeter.
Events containing any jets with $\pt > \gev{50}$ overlapping dead tile modules are rejected if the azimuthal angular separation between the \MET and the jet, $\Delta \phi(\MET, \text{jet})$, is less than 0.3.
All SR events must meet these common event-level selection criteria.

A CR is populated by events triggered by a single photon with $\ptPhoton > \gev{140}$.
These events must pass the same common selection criteria as above, \textit{except} that they must fail the combined \met trigger and offline \met selections, so that the CR is orthogonal to the SRs.
This region was chosen as the CR because the events triggered by a single photon ($\gamma$ + jets) and \met (\Zjets) have a similar topology except for the difference between photon and \met, and they have large statistics and negligible signal contamination from models of interest.


%
%
%

%
\subsection{Signal region definitions}
\label{sec:sr_defs}

This section defines the two SRs for this search: SDV SR and FDV SR, based on the vertexing algorithm.
The FDV SR is further divided into two mutually exclusive sub-regions: the single fuzzy vertex sub-region (1FDV) for models with exactly one LLP decay, and the two fuzzy vertices sub-region (2FDV) for models involving pair production of LLPs decaying to two or more distinct FDVs.
These sub-regions are analyzed independently, and their results are statistically combined for the final FDV SR result.
For FDV SR and its associated CR, an additional event-level selection called \textit{strict NCB veto} is applied to reject the NCB processes more strictly.
It rejects the events if the fraction of the charged component of the leading jet is lower than 10\% of the fraction of energy that is found in its single most energetic calorimeter layer.
The strict NCB veto filters out NCB processes to a negligible level, but also removes signal events where a jet from an LLP becomes the leading jet.
Therefore, it is only used for FDV SR, which targets models where the leading jet generally comes from initial-state radiation.

After common event-level selections, candidate DVs must pass baseline criteria (\Tab{\ref{tab:baseline_vertex_selections}}) to ensure they are well-measured and reduce background.
The reconstructed vertex position must be in the fiducial region, which is defined as a radial displacement from the beamline axis $\rdv < 300$~mm and axial displacement from the origin $|z_\mathrm{DV}| < 300$~mm, where $x_\mathrm{DV}, y_\mathrm{DV}$, and $z_\mathrm{DV}$ are the cartesian coordinates of the DV.
The requirement that the vertex is separated in the transverse plane from all reconstructed CVs, \rxyWrtCV, is to avoid mislabeling CVs as secondary vertices and suppress the background from the decay of heavy-flavor hadrons.
It is also required that the vertex is not reconstructed near an energetic photon, \deltaRgamma, to remove DVs arising from photon conversions.
The selection on the maximum track transverse momentum (\maxpTtrack) with respect to the scalar sum of the transverse momentum of all tracks in the vertex is in place to suppress background from vertices that contain mismeasured tracks.
SDVs must have a reduced $\chi^2$ in the vertex fitting, \redChiSq, of less than 5 to remove the misreconstructed vertices.

\begin{table}[h!]
\centering
\caption{Baseline vertex selection criteria for SDV SR and FDV SR (including the 1FDV and 2FDV subregions). \label{tab:baseline_vertex_selections}}
\resizebox{0.9\columnwidth}{!}{
\def\arraystretch{1.2}
\begin{threeparttable}
\begin{tabular}{c||c|c}
\toprule
Baseline vertex selections & SDV SR & FDV SR \\
\midrule
Vertexing algorithm & Standard DV algorithm (SDV) & Fuzzy DV algorithm (FDV)  \\
Vertex position & \multicolumn{2}{c}{$\rdv < 300$ mm, \quad $|\zdv| < 300$ mm} \\
\rxyWrtCV & \multicolumn{2}{c}{$ > 4$ mm for any CV} \\
Material map veto  & \multicolumn{2}{c}{Strict veto required (\Sect{\ref{sec:material_map}})} \\
\deltaRgamma  & \multicolumn{2}{c}{$> 0.1$, for photons with $\pT(\gamma) > \gev{60}$} \\
\maxpTtrack/(scalar sum $p_\mathrm{T}^{\text{track}}$) & $ < 0.95$ & $ < 0.9$ \\
\redChiSq & $ < 5$ & -- \\
\bottomrule
\end{tabular}%
\end{threeparttable}
}
\end{table}

Vertices passing the baseline vertex selections are used to develop and validate the background estimation methods for the SRs.
Additional quality selections are applied to tracks associated with the DVs in order to remove tracks that appear to originate from background processes from DVs.
These selections are placed on track variables including the track $d_{0}$-significance, the azimuthal angle between the track and the PV-SDV vector \dPhiPVSDV and the PV-FDV vector \dPhiPVFDV, and the three-dimensional angle between the track and the PV-DV vector $\alpha$, which are designed to reject prompt tracks originating from the CVs.
The DV properties, including \nDVTracks, \mDV, and \nDVSeeds, are recalculated after removing tracks failing the quality selections, while the vertex refitting is not performed and the vertex position is not redefined.
The \textit{signal vertex selections} are then applied to generate the SRs, summarized in \Tab{\ref{tab:SRsignalSelections}}.
The signal vertex selections are applied to the following values: the track multiplicity, \nDVTracks, the invariant mass of the DV, \mDV, the number of selected tracks, \nDVSelTracks, the separation between the DV and closest jet, \dRJetDV, the number of two-track seed vertices associated with the FDV, \nDVSeeds, the angular separation in $\eta$ direction from the PV to the DV, $|\dEtaPVDV|$, and the maximal $\Delta \eta$ between a single track and the sum of the four-vector of the other tracks, \dEtaTracks.
The requirement placed on \nDVSeeds for the FDV sub-regions ensures a high degree of compatibility among the tracks.
It is required that the number of vertices satisfying these criteria, \NDV, for the SDV and 1FDV region is at least one, whereas the 2FDV region requires two or more.
For the 1FDV sub-region it is also required that there is no additional FDV that satisfies the full SR selections for the 2FDV sub-region in order to keep the 1FDV and 2FDV sub-regions mutually exclusive.

\begin{table}[h!]
\centering
\caption{Full vertex selection criteria for SDV SR and FDV SR (including the 1FDV and 2FDV subregions). \label{tab:SRsignalSelections}}
\resizebox{0.9\columnwidth}{!}{
\def\arraystretch{1.2}
\begin{threeparttable}[h]
\begin{tabular}{c||ccc}
\toprule
\multirow{2}{*}{Full SR vertex selection} & \multirow{2}{*}{SDV SR} & \multicolumn{2}{c}{\quad \quad \quad \quad FDV SR} \\
& &  1FDV & 2FDV  \\
\midrule
Vertices & SDV & \multicolumn{2}{c}{\quad \quad \quad \quad FDV} \\
\nDVTracks & $\geq$ 5 & $\geq$ 5 & $\geq$ 4  \\
\mDV~[\gevonly]& $>$ 10 & $>$ 10 & $>$ 1.5  \\
\nDVSelTracks & $\geq 2$ & -- & --\\
\nDVSeeds & -- & $\geq \binom{\nDVTracks}{2} - 1$ & $\geq$ \nDVTracks  \\
\dRJetDV & --  & -- & $< 0.4$  \\
|\dEtaPVDV| & -- & -- & $< 2.5$  \\
\dEtaTracks & -- & < 3.5 & -- \\
\NDV & $\geq 1$  & $ \geq 1\tnote{$\dagger$}$ & $ \geq 2$ \\
\bottomrule
\end{tabular}%
\begin{tablenotes}[para,flushleft,online,normal] %
\item[$\dagger$] The event must not have any additional FDVs satisfying the 2FDV selections.
\end{tablenotes}
\end{threeparttable}
}
\end{table}

The signal efficiencies are key metrics used in optimizing selection criteria and validating this analysis.
\Tab{\ref{tab:sigeff_table}} summarizes the signal efficiency values for a range of
representative sample signal points, highlighting the spread of efficiency values across
samples in their respective SRs. Note that for the \HiggsPortalModel model, efficiency is
defined relative to the signal samples used at a given lifetime. The lower efficiency
at $\tau = \ns{3.3}$ is because \ggH production is only included at that lifetime.

\begin{table}[h!]
\renewcommand{\arraystretch}{1.2}
\centering
\caption{Summary of signal efficiencies in each signal region for models with fixed mass parameters and varied lifetimes.\label{tab:sigeff_table}}
\begin{tabular}{c||c c c c}
\toprule
& & \multicolumn{3}{c}{\textbf{Efficiency [\%]}} \\
\textbf{Model} & \textbf{Lifetime $\tau$ [ns]} & \textbf{SDV SR} & \textbf{1FDV SR} & \textbf{2FDV SR} \\ \midrule

\multirow{3}{*}{\parbox[c]{4cm}{\centering \RhadronModel \\ \mgluino = \gev{1600} \\ \mneutralino = \gev{100}}}
& 0.01 & 6.4 & 0.29 & 0.084 \\
& 0.1  & 41  & 1.1  & 0.36 \\
& 3.0  & 6.9 & 0.09 & 0.0029 \\ \midrule

\multirow{3}{*}{\parbox[c]{4cm}{\centering \AxinoModel ($ZZ$) \\ \maxino = \gev{400} \\ \mneutralino = \gev{800}}}
& 0.0045 & 5.15  & 0.99 & 0.11 \\
& 0.097  & 40.19 & 4.1  & 0.30 \\
& 1.5    & 9.26  & 0.48 & 0.012 \\ \midrule

\multirow{3}{*}{\parbox[c]{4cm}{\centering \BinoWinoModel \\ \mneutralino = \gev{650} \\ \mslneutralino = \gev{700}}}
& 0.03 & 0.13  &  0.37 & 0.055 \\
& 0.3  & 0.10  & 0.20  & 0.024 \\
& 3.0  & 0.014 & 0.037 & 0.006 \\  \midrule

\multirow{3}{*}{\parbox[c]{4cm}{\centering \HiggsPortalModel \\ \mS = \gev{55}}}
& 0.0033 & \num[round-mode=places, round-precision=3]{0.006907114379232958} & \num[round-mode=places, round-precision=3]{0.0073752006773727275} & \num[round-mode=places, round-precision=3]{0.021520830186410718} \\
& 0.033  & \num[round-mode=places, round-precision=3]{0.026224882238445178} & \num[round-mode=places, round-precision=3]{0.02291186666936838} & \num[round-mode=places, round-precision=3]{0.057389437859013376} \\
& 3.3    & \num[round-mode=places, round-precision=4]{0.0003199622944} & \num[round-mode=places, round-precision=4]{0.0003048249214} & \num[round-mode=places, round-precision=4]{0.0001756155031}\\ \midrule
\end{tabular}
\end{table}

For an accurate estimate of the background contributions to each SR, a data-driven approach is employed.
The background estimates for each SR are independently derived and optimized accordingly.
The following sections detail the background estimation methods and their validation per SR.


%
%
%

%
\subsection{Background estimates and validation regions}
\label{sec:validation_regions}

Due to the origin and composition of the expected background contributions, multiple VRs are defined for each SR. These VRs are chosen to be kinematically similar to the SRs while being statistically independent, allowing us to test the background prediction techniques in adjacent phase spaces.
The VRs are constructed by modifying key selection criteria such as \nDVTracks, \mDV, \nDVSeeds, and whether the material map veto flags them as being positioned either inside or outside or detector material.
For the SDV SR and 1FDV SR, VRs are defined by selecting vertices with exactly four tracks in the middle mass range (10--20 GeV), or by selecting vertices in the lower mass ranges (5--10 GeV) with more than four tracks. The photon-triggered CRs are defined with similar boundaries.
The VR definition for the 2FDV SR is discussed in \Sect{\ref{sec:2FDVValidation}}.
The background estimation methods are tested in each of these VRs to ensure reliable predictions in the corresponding SRs.
These VRs are built so that the amount of potential signal contamination from any considered model is not significant.
The maximum expected signal contamination is well below 30\% for the non-excluded parameter space.
Since it is confirmed that the signal is expected to produce a much larger excess in the SR than the VRs, this level of contamination does not compromise the validation procedure.
%

%
%
%
%
%
%
%
%
%
%
%
%
%
%
%
%
%
%
%
%
%
%
%
%
%
%

%
\subsubsection{SDV background estimate}
\label{sec:vsi_bkg}

Searches for SDVs with high invariant mass and multiple tracks face minimal background due to the rarity of SM processes producing such signatures.
Background SDVs in general primarily arise from:
\begin{itemize}
\item \textbf{Hadronic interactions (HI)}, where particles interact with detector material, producing collimated tracks and low-mass vertices.
\item \textbf{Accidental crossings (AX)}, where unrelated tracks from the primary vertex or pileup attach to a SDV\footnote{Note that the lower-mass or lower-multiplicity SDVs that gives rise to the MV and AX backgrounds can come from a variety of potential sources, including boosted heavy-flavor decays or from HI. The origin of these SDVs is not directly taken into account during the background estimate procedure.}, increasing its apparent mass and track count.
\item \textbf{Merged vertices (MV)}, where separate SDVs are mistakenly combined into a single vertex with higher mass and multiplicity.
\end{itemize}

All three sources of background are estimated inclusively with a data-driven technique that exploits the relationship between SDVs and the track density of the event to predict the number of background vertices.
The estimated number of background events in the SR, $N_\text{bkg}$, is computed using an event-DV probability (EDP) $P(\textrm{vertex}|\nTracksEvent, \nBTag)$, where \nTracksEvent is the total number of tracks associated with any CV and \nBTag is the number of $b$-tagged jet count per event.
Defined in \Eqn{\ref{eq:edp}}, the EDP represents the probability of reconstructing an SR-like vertex satisfying the baseline vertex selections with the additional requirement: $\mdv > 5~\text{GeV}$ and $\nDVTracks \geq 4$, which is measured in the CR.
To correct for the effects of differences in the distributions of the number of pileup collisions, the events in the CR are weighted such that the distributions of the number of pileup collisions match those of the events passing the common selections (\Sect{\ref{sec:eventlevel}}) when calculating the EDP.
This selection on \mdv and \nDVTracks is intentionally less stringent than the SDV SR requirement, as the stricter criteria result in too few events to generate a non-trivial estimate.
The extrapolation factor $f$, obtained from the CR events according to \Eqn{\ref{eq:fFactor}}, corrects for this by extrapolating to the SR selection values.
The dependency of $f$ on \nTracksEvent{} and \nBTag is confirmed to be negligible compared to the other uncertainties in the background estimate.
The EDP is applied to all events passing the common event-level selections described in \Sect{\ref{sec:eventlevel}} for the SR region, and the background estimate is then computed as in \Eqn{\ref{eq:nBkg}}, where the sum is taken over the SR.
This SDV background estimate method (also used for 1FDV) counts number of DVs instead of events with DVs. Since events with more than one DV with \nDVTracks $\geq 4$ and $\mDV > 5$~GeV are very rare, the uncertainty from using this approximation is small compared to other uncertainties in the background estimate.
\begin{equation}
\label{eq:edp}
P\textrm{(vertex }|\nTracksEvent, \nBTag)  = \frac{N_{\mathrm{Vertex}}^{\mathrm{CR}}(\nTracksEvent, \nBTag)}{N_{\mathrm{Events}}^{\mathrm{CR}}(\nTracksEvent, \nBTag)}
\end{equation}
\begin{equation}
\label{eq:fFactor}
f = \frac{N_{\mathrm{vertex}}^{\mathrm{CR}}(\mdv > \gev{10}, \nDVTracks \geq 5)}{N_{\mathrm{vertex}}^{\mathrm{CR}}(\mdv > \gev{5}, \nDVTracks \geq 4)}
\end{equation}
\begin{equation}
\label{eq:nBkg}
N_\text{bkg} = f \cdot \sum_{\mathrm{events}} P\textrm{(vertex}|\nTracksEvent, \nBTag)_{\mathit{i}}
\end{equation}

This method predicts an expected background of $0.6 \pm 0.4$ events in the SDV SR, with the dominant uncertainty being statistical.

\subsubsection{SDV background estimate validation}
\label{sec:vsiValidation}
\begin{figure}[!htb]
\centering
\includegraphics[width=0.7\textwidth]{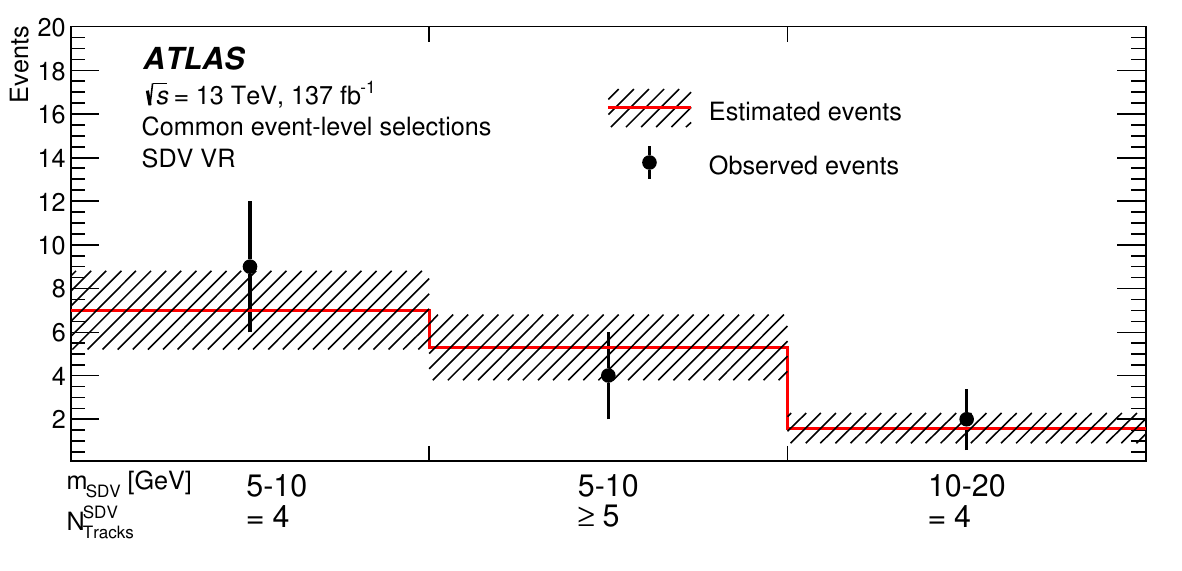}
\caption{
Estimated and observed events in the sideband VRs of the SDV SR.
The gray hashed band shows the background estimate uncertainties from the CR's statistics and from the distributions of the number of pileup collisions between events in the CR and the events that pass common event-level selections.
}
\label{fig:vsiValidation}
\end{figure}

The predictions of the background estimation method are tested in three statistically independent sideband VRs.
Events in these sideband VRs are required to satisfy all the SDV SR requirements, except for the relaxed conditions of $5 < \mSDV < \gev{10}$ or $\mSDV < \gev{20}$ with $\nTracks = 4$.
The full background estimate procedure, including calculating an EDP and extrapolation factor $f$ is done for each VR.
The details of the VRs and a comparison of the predicted and observed number of events in each VR are shown in \Fig{\ref{fig:vsiValidation}}.
The predictions of the background estimate are found to agree with the number of observed events in data for all three of the VRs within the uncertainty.

In addition to validating the background estimate in sideband regions, a separate background estimate is performed where the contribution from each source -- hadronic interactions, merged vertices and accidental crossings -- is estimated independently.
Each contribution is estimated using data-driven techniques that construct mass templates describing the shape of \mSDV for background SDVs from the particular background source.
These methods are described and validated in previous ATLAS searches for displaced vertices~\cite{SUSY-2016-08, SUSY-2018-13}, and will only be briefly detailed here:
\begin{itemize}
\item \textbf{HI}:
The HI SDV mass distribution is modeled as
\begin{equation*}
N_\text{SDV} (\mdv;b,C,l) = \left[ \frac{1}{C(\mdv-b)} + \exp\left(\frac{\mdv-b}{l}\right) \right]
\end{equation*}
where $b$, $C$, and $l$ are fitted constants, and the resulting fit is integrated to produce an estimate of the number of HI background SDVs created.
\item \textbf{AX}:
The rate of accidental crossings is estimated from data, using candidate $2$- and $3$-track $\kso \rightarrow \pi^+ \pi^-$ decay SDVs, with the identified AX tracks' kinematics saved.
Artificial accidentally-crossed SDVs are created by attaching the above AX tracks to data SDVs.
Their mass distributions are scaled by the accidental crossing rate, and integrated to yield an estimated number of background SDVs.
\item \textbf{MV}:
Merged SDVs are created by artificially overlapping events in data and reclustering vertices.
The resulting vertices are used to estimate the merging rate -- by comparing distances between vertices to those in the original data -- and to produce MV mass templates.
The templates are rescaled by the merging rate, and integrated to yield an estimated number of background SDVs.
\end{itemize}
The total expected background obtained in each VR and the SDV SR agree within statistical uncertainty with the expected background from the inclusive method.

The estimated number of background events in SDV SR with its uncertainty is shown in Table~\ref{tab:uncertaintySummary}.



%
\subsubsection{1FDV background estimate}
\label{sec:1FuzzyInclusive}

The background estimate for 1FDV SR uses the same approach as the data-driven SDV SR background estimate, described in \Sect{\ref{sec:vsi_bkg}}, using the FDVs instead of SDVs.
The expected background is predicted to be
$0.8 \pm 0.5$ events in the 1FDV SR.
The dominant uncertainty comes from the statistical uncertainty in the number of FDVs in the CR.
The uncertainty also includes the effect of the difference in the distributions of the number of pileup collisions between the photon-triggered CR and \met triggered regions. %

\subsubsection{1FDV background estimate validation}
\label{sec:1FDVValidation}

\begin{figure}
\centering
\includegraphics[width=\textwidth]{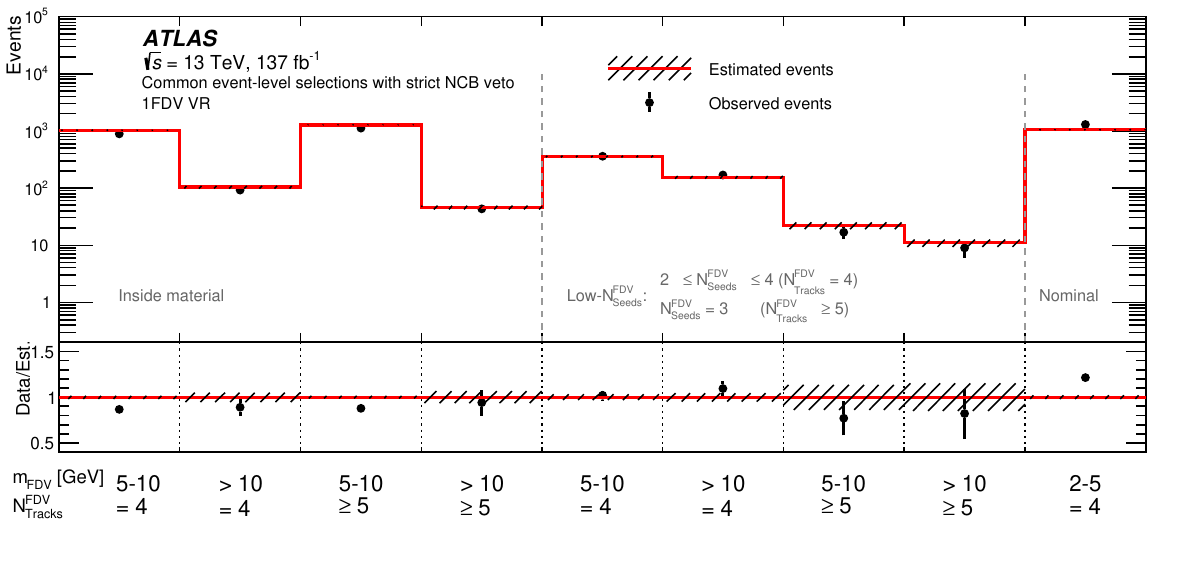}
\caption{Estimated and observed events in the sideband VRs of the 1FDV VRs.
The gray hashed band shows the background estimate uncertainties from the CR's statistics and from the distributions of the number of pileup collisions between events in the CR and the events that pass common event-level selections and the strict NCB veto. The non-closure uncertainty discussed in the text is not included here.
}
\label{fig:1FDVValidation}
\end{figure}

Events in the VR are required to meet the same criteria as the 1FDV SR, except that the \nFDVTracks requirements are relaxed from $\nFDVTracks \geq5$ to $\nFDVTracks \geq4$, and the \mFDV requirements are changed from $\mFDV > 10~\mathrm{GeV}$ to $2 < \mFDV < 5~\mathrm{GeV}$, which defines the nominal VR.
Another set of sideband VRs is developed by loosening the conditions on \nFDVSeeds.
To validate the effects of background hadronic interactions (\Sect{\ref{sec:material_map}}), a third VR is defined by requiring vertices to be located inside the material determined by the material map veto flags.
The selection criteria for these VRs are summarized in \Fig{\ref{fig:1FDVValidation}}.
For each VR, the full background estimation procedure described in \Sect{\ref{sec:vsi_bkg}} is applied, and the predicted and observed numbers of FDVs are compared.
\Fig{\ref{fig:1FDVValidation}} shows the estimated and observed event yields in the 1FDV VRs, which roughly agree but are not statistically compatible across all regions.
This discrepancy indicates potential non-closure effects, for example subtle differences in jet fragmentation or soft hadronic activity between the photon and \met triggered regions.
To account for these effects, a 22\% relative uncertainty, which corresponds to the maximum discrepancy between the estimated and observed event yields for each VR bin, is applied as a non-closure systematic uncertainty for the 1FDV background estimate.
The estimated number of background events in 1FDV SR with its uncertainty is shown in Table \ref{tab:uncertaintySummary}.

%
\subsubsection{2FDV background estimate}
\label{sec:2FuzzyInclusive}

The background estimation strategy for the 2FDV SR follows a similar inclusive approach to the SDV and 1FDV SRs, leveraging correlations between FDVs and jet properties rather than track density.
This estimate is obtained using a probability distribution over the jets in the \met triggered region, defined by the probability of an individual event containing a FDV near a jet.
This probability is parameterized based on the jet's \pt and whether or not the jet is $b$-tagged.
The estimates for the SDV, 1FDV, and 2FDV regions were initially all tested taking into account both the event- and jet-DV probabilities. It was found that the heavier SDV and 1FDV regions were primarily dependent on the event-DV correlations, while the less selective 2FDV region were primarily dependent on the jet-DV correlations due to differing kinematics.
As the statistical uncertainty originating from having the estimate divided up into too many probability bins is larger than the contribution of the subdominant correlation, only the primary correlation is used.
The individual jet probabilities, referred to as the jet-FDV probability (JFP), are calculated in the photon-triggered CR as shown
in \Eqn{\ref{eq:JFP}}, where $N_{n\mathrm{ FDVs}}^{\mathrm{CR}}(b_{\mathrm{tag}}, \pt{} )$ represents the number of instances where $n$ FDVs are matched to jets of $(b_{\mathrm{tag}}, \pt{} )$, and $N_{\mathrm{Jets}}^{\mathrm{CR}}(b_{\mathrm{tag}}, \pt{})$ represents the total number of jets that exist with $(b_{\mathrm{tag}}, \pt{} )$.

\begin{equation}
\label{eq:JFP}
\mathrm{JFP}_{(n)}(\mathrm{FDV} ~|~ \mathrm{jet}(b_{\mathrm{tag}},\pt{})) = \frac{N_{n\mathrm{ FDVs}}^{\mathrm{CR}}(b_{\mathrm{tag}}, \pt{})}{N_{\mathrm{Jets}}^{\mathrm{CR}}(b_{\mathrm{tag}}, \pt{})}
\end{equation}

In the CR, the JFP is determined using jets spatially matched to FDVs within $\Delta R \leq 0.4$.
If multiple FDVs are matched to the same jet, it contributes to the estimation of \JFP{\geq 2}, whereas if a jet only matches a single FDV it contributes to \JFP{1}.

The number of background events with two or more FDVs, $N_{\text{bkg}}^{\geq 2 \, \text{FDV}}$ is then calculated in
\Eqn{\ref{eq:nBkg_2FDV}}, with the number of background events with $n$ background FDVs, $N_{\text{bkg}}^{n \, \text{FDV}}$, given by
\Eqns{\ref{eq:nBKG_2FDV_0}}{\ref{eq:nBKG_2FDV_1}} below.
\begin{align}
N_{\text{bkg}}^{\geq 2 \, \text{FDV}} &= N_{\text{bkg}} - N_{\text{bkg}}^{1 \, \text{FDV}} - N_{\text{bkg}}^{0 \, \text{FDV}} \label{eq:nBkg_2FDV} \\
N_{\text{bkg}}^{0 \, \text{FDV}} &= \sum_{\text{events }E}  \left[ \prod_{\text{jets }j\text{ in event }E} \left( 1 - \text{JFP}_{(1)} - \text{JFP}_{(\geq 2)} \right) \right] \label{eq:nBKG_2FDV_0} \\
N_{\text{bkg}}^{1 \, \text{FDV}} &= \sum_{\text{events }E}  \left[ \prod_{\text{jets }j_E\text{ in event }E} \text{JFP}_{(1)}
\left( \prod_{\text{jets }j\neq j_E\text{ in event }E} \left( 1 - \text{JFP}_{(1)} - \text{JFP}_{(\geq 2)} \right) \right) \right] \label{eq:nBKG_2FDV_1}
\end{align}
An expected background of $1.3 \pm 0.6$ events is predicted in the 2FDV SR, with the dominant uncertainty being statistical.

\subsubsection{2FDV background estimate validation}
\label{sec:2FDVValidation}

Events in the 2FDV VR are required to satisfy the 2FDV SR conditions, except for the relaxed requirement of \nFDVTracks $= 3$ or $2 \leq \nFDVSeeds < \nFDVTracks$, creating sideband VRs.

\begin{figure}
\centering
\includegraphics[width=\textwidth]{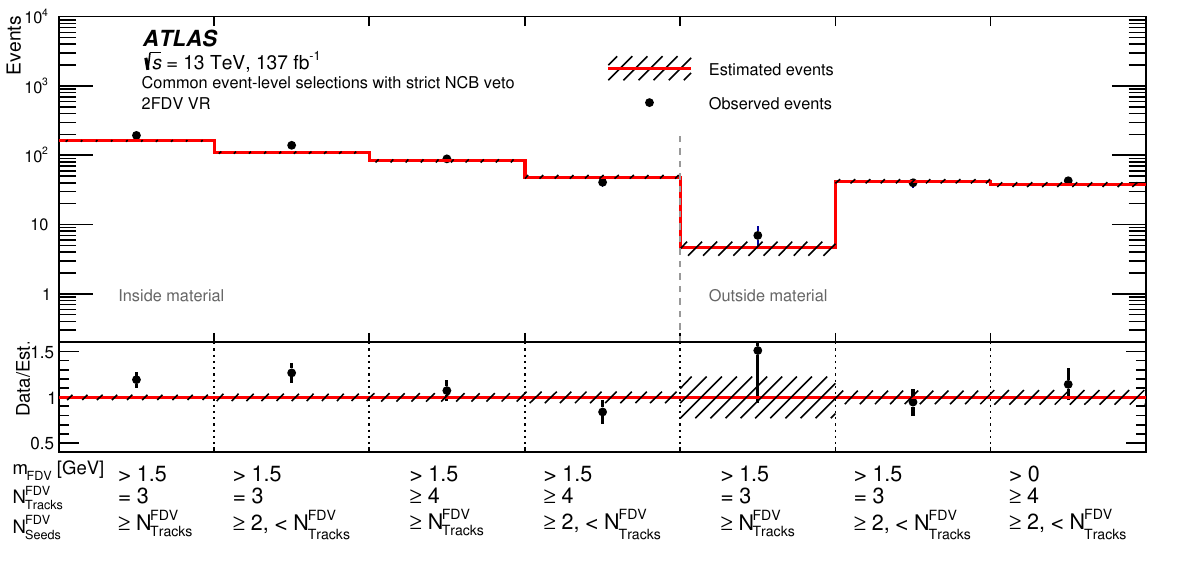}
\caption{
Estimated and observed number of events in the sideband VRs of the 2FDV SR.
The gray hashed band shows the relative background estimate uncertainties from the CR's statistics and distributions of the number of pileup collisions between events in the CR and the events that pass common event-level selections and the strict NCB veto. The non-closure uncertainty discussed in the text is not included here.
}
\label{fig:2FDVValidation}
\end{figure}

In addition to the \nFDVTracks and \nFDVSeeds - sideband VRs, the estimate is validated using both the strict inside material and strict outside material map vetoes defined in \Sect{\ref{sec:material_map}}.
Exceptionally, the outside-material, greater-than-three-tracks, and low-seed validation regions were loosened to increase statistics by relaxing the material map veto and including the entire mass range (\mFDV$>0$~GeV).
The selection criteria for these two VRs are shown in \Fig{\ref{fig:2FDVValidation}}.
The $N^{\geq 2 \, \text{FDV}}_{\text{bkg events}}$ is calculated for each VR using the background estimation method outlined in \Sect{\ref{sec:2FuzzyInclusive}}.
\Fig{\ref{fig:2FDVValidation}} compares each VR's predicted and observed number of events with two or more FDVs.
As with 1FDV, the estimated and observed number of events are roughly consistent, but the discrepancy between them could not be explained by statistical fluctuation alone, indicating potential non-closure effects.
Better closure is more important for VRs that differ from the SR by only one selection criterion (the 3rd, 5th, and 7th bins in \Fig{\ref{fig:2FDVValidation}}) than those that differ by two (1st, 4th, and 6th bins) or more (2nd) selection criteria.
To account for potential non-closure effects, a 20\% relative uncertainty is applied as a non-closure systematic for the 2FDV background estimate.

The estimated number of background events in 2FDV SR with its uncertainty is shown in Table \ref{tab:uncertaintySummary}.




%
\section{Systematic uncertainties}
\label{sec:systematics}

The uncertainties considered in this analysis can be divided into three categories.
The first are uncertainties arising from the background estimation methods, which are detailed in Section~\ref{sec:bkg_syst}.
The second are uncertainties from object reconstruction in the detector, which are detailed in Section~\ref{sec:reco_syst}.
Finally, theory uncertainties from the signal models must also be considered in the interpretation of results, which are detailed in Section~\ref{sec:theory_syst}.

%

\subsection{Background estimate uncertainties}
\label{sec:bkg_syst}

Multiple sources contribute to background estimate uncertainties and are evaluated for each SR.
The dominant source in each region is statistical, which is propagated to the final estimate using a pseudo-experiment approach (similar to how the final fit is performed in Section \ref{sec:fitting}). The standard deviation of the estimated background from 10,000 pseudo-experiments is then assigned as an uncertainty on the final background estimate. In the FDV regions, an additional non-closure uncertainty of 22\% for the 1FDV SR and 20\% for the 2FDV SR is
assigned to cover discrepancies between this estimate and observed data in some of the VRs.

A further uncertainty is assigned to account for differences in the distributions of the number of pileup collisions between events in the CRs (where the event- or jet-DV probabilities are calculated) and the SRs (where the event- or jet-DV probabilities are applied).
The estimated background using the pileup weighted events is taken as the central value of the background estimate, as described in Section \ref{sec:vsi_bkg}.
To assess this uncertainty, the estimated background is re-evaluated using unweighted events.
The difference between the weighted and unweighted estimates is assigned as an uncertainty on the background estimate.
Like the statistical uncertainty, the pileup uncertainty is evaluated for the background estimate in each region.
A maximum pileup uncertainty of 5\% is assessed in the SDV SR. Each of these uncertainties is summarized in Table~\ref{tab:uncertaintySummary}.

\begin{table}[]
\centering
\caption{A summary of the breakdown of background estimate uncertainties per signal region. \label{tab:uncertaintySummary}}
\begin{tabular}{ c | c | c c c | c}
\toprule
Region & Estimated  & \multicolumn{3}{c|}{Uncertainties} & Total uncertainty \\

& background & Statistical & Pileup & Non-closure \\
\midrule
SDV SR & $0.6 \pm 0.4$ & $\pm 70 \%$ & $\pm 5 \%$ & -- & $\pm 71 \%$ \\
\midrule
1FDV SR & $0.8 \pm 0.5$ & $\pm 57 \%$ & $\pm 2 \%$ & $\pm 22 \%$ & $\pm 61 \%$ \\
2FDV SR & $1.3 \pm 0.6$ & $\pm 45 \%$ & $\pm 4 \%$ & $\pm 20 \%$ & $\pm 50 \%$ \\
\bottomrule
\end{tabular}
\end{table}


%
\subsection{Reconstruction uncertainties}
\label{sec:reco_syst}

Potential sources of reconstruction uncertainty -- those stemming from how physics objects are reconstructed -- are the mismeasurement of offline and online \met, as well as track reconstruction. Their effects are estimated and propagated to the predicted MC signal yields.
Offline \met uncertainties are estimated by rescaling the different contributions to the measured \met (e.g. the soft term, see Section~\ref{sec:objects}), via their associated systematic uncertainties~\cite{JETM-2020-03}.
Sources include the jet energy scale (JES) and resolution (JER)~\cite{JETM-2018-05} (up to 1.9\%) and uncertainties in the scale and resolution of the soft term (up to 0.4\%)~\cite{JETM-2020-03}.
For online \met, uncertainties are estimated by deriving scale factors to account for inefficiency in data not modeled in MC, in the $150$ -- \gev{180} range.

\begin{figure}[!htb]
\centering
\subfloat[\label{fig:tk_2016_sdv_ratio}]{\includegraphics[width=0.33\textwidth]{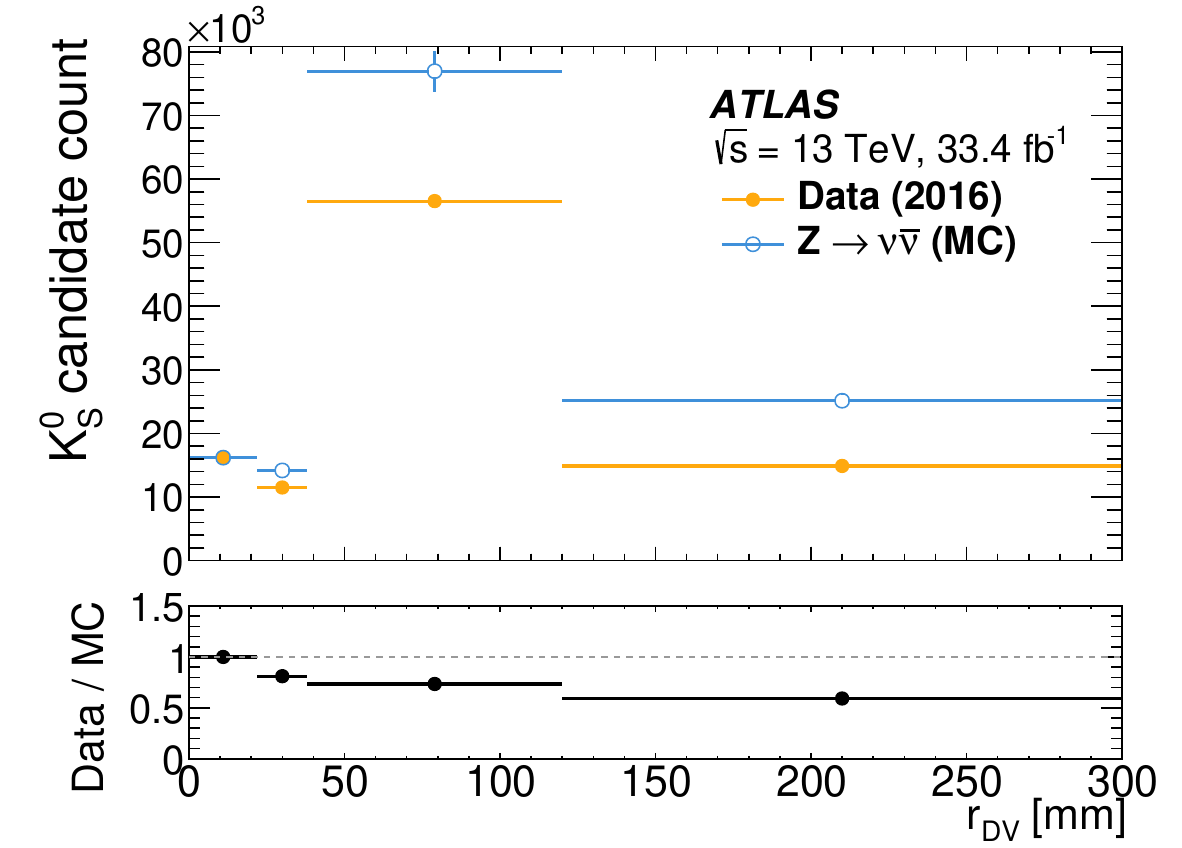}}
\subfloat[\label{fig:tk_2017_sdv_ratio}]{\includegraphics[width=0.33\textwidth]{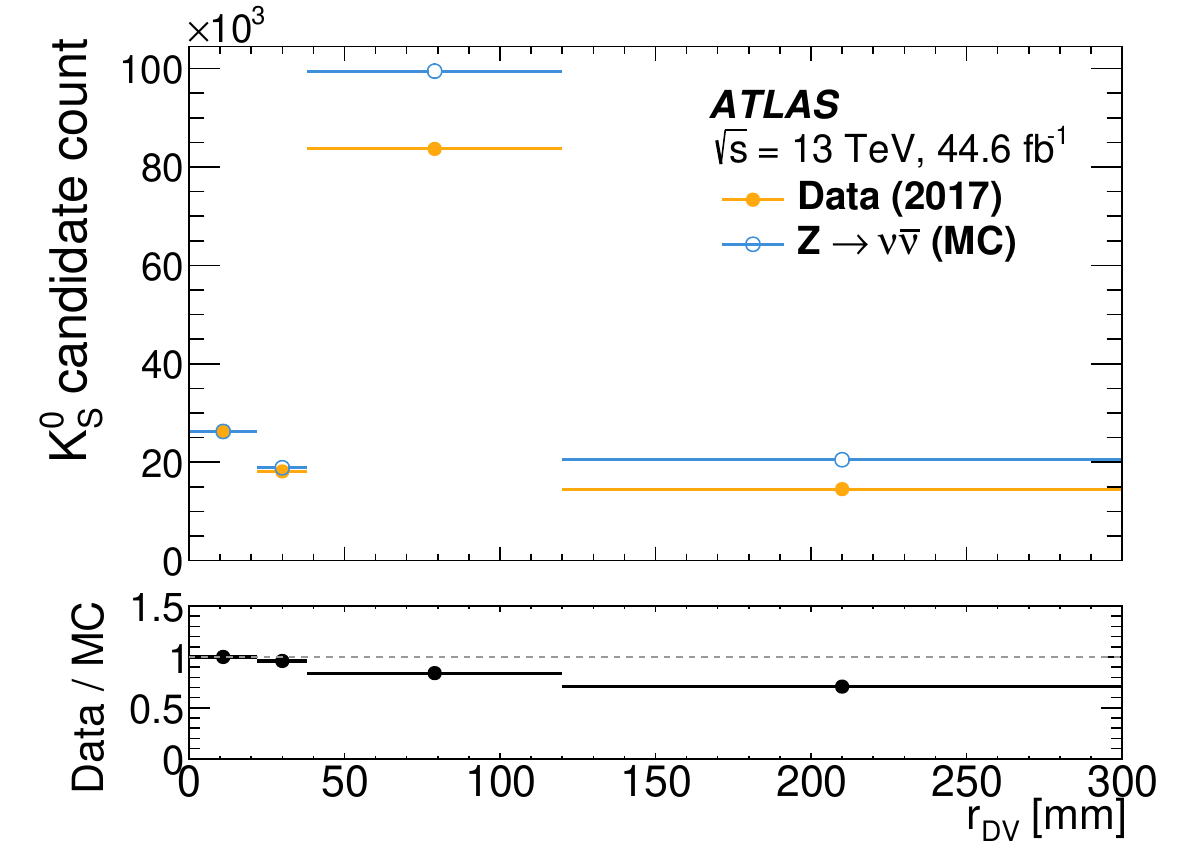}}
\subfloat[\label{fig:tk_2018_sdv_ratio}]{\includegraphics[width=0.33\textwidth]{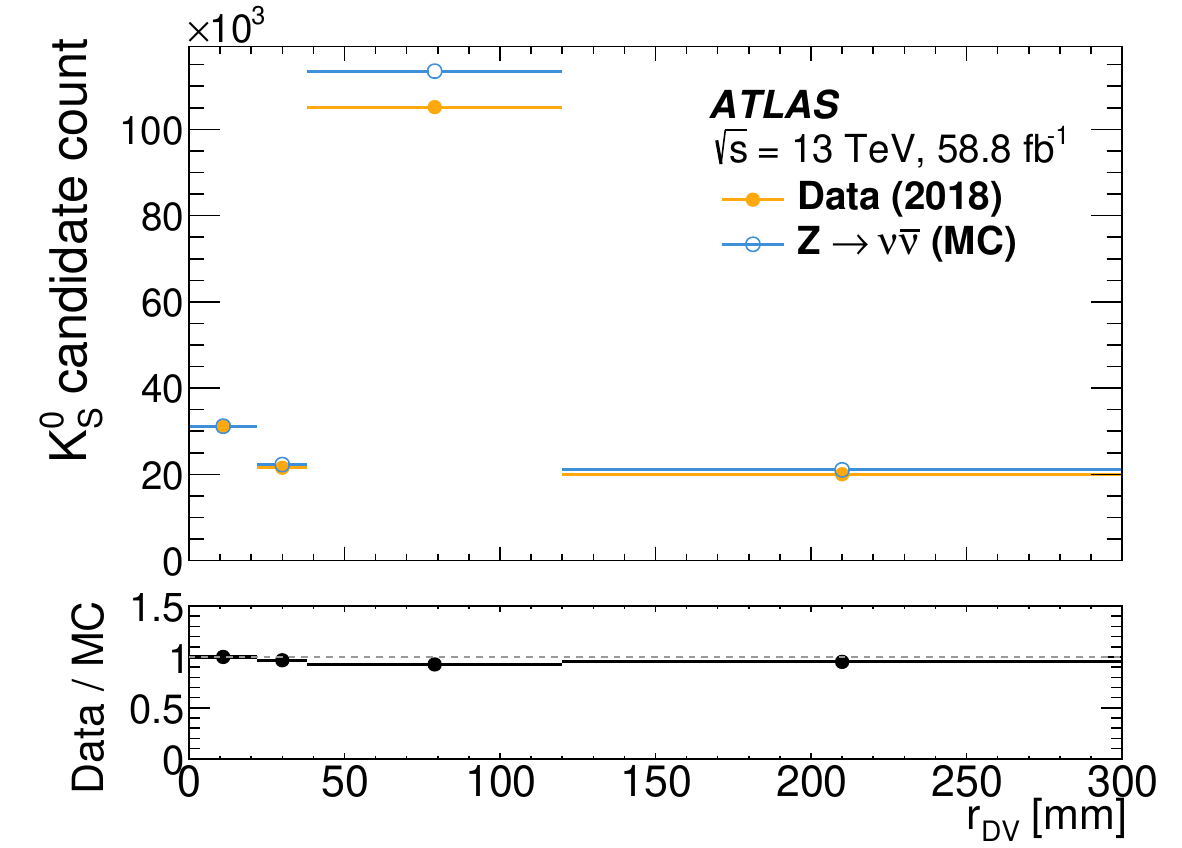}} \\
\subfloat[\label{fig:tk_2016_sdv_unc}]{\includegraphics[width=0.33\textwidth]{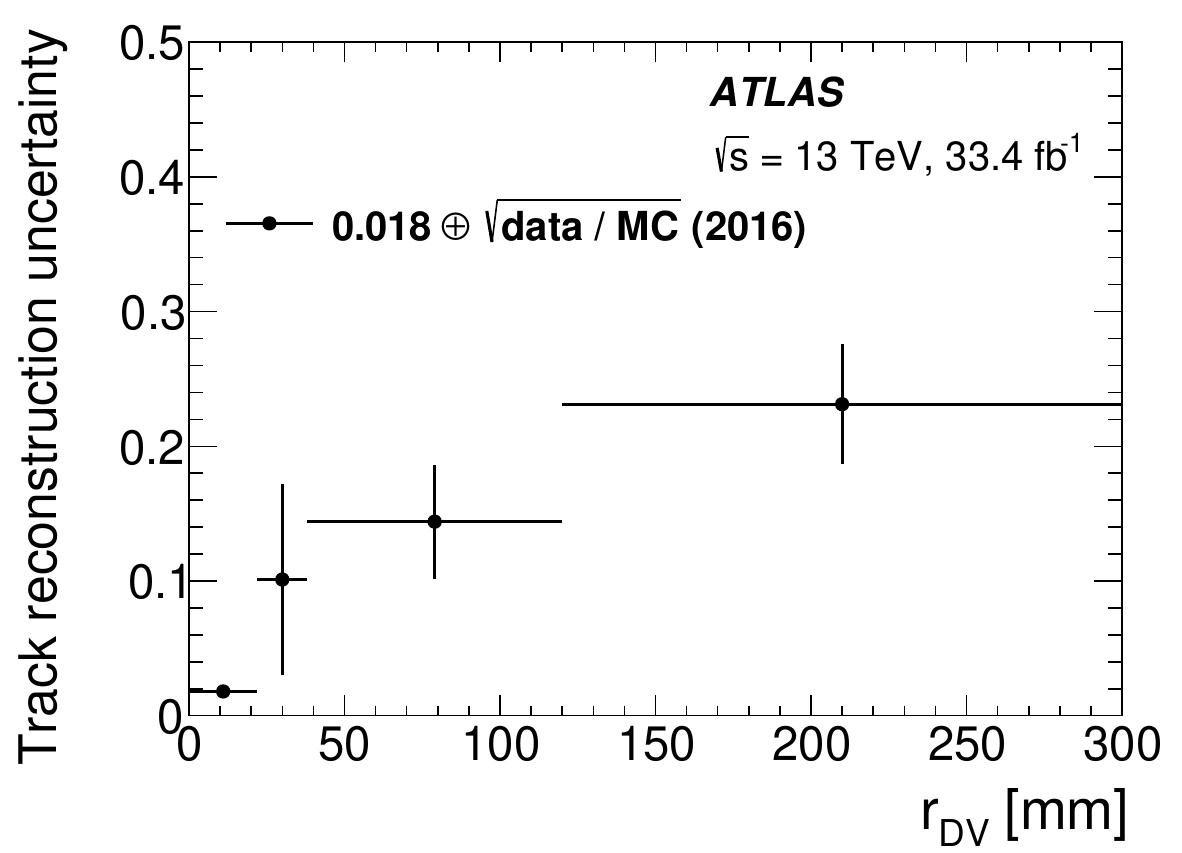}}
\subfloat[\label{fig:tk_2017_sdv_unc}]{\includegraphics[width=0.33\textwidth]{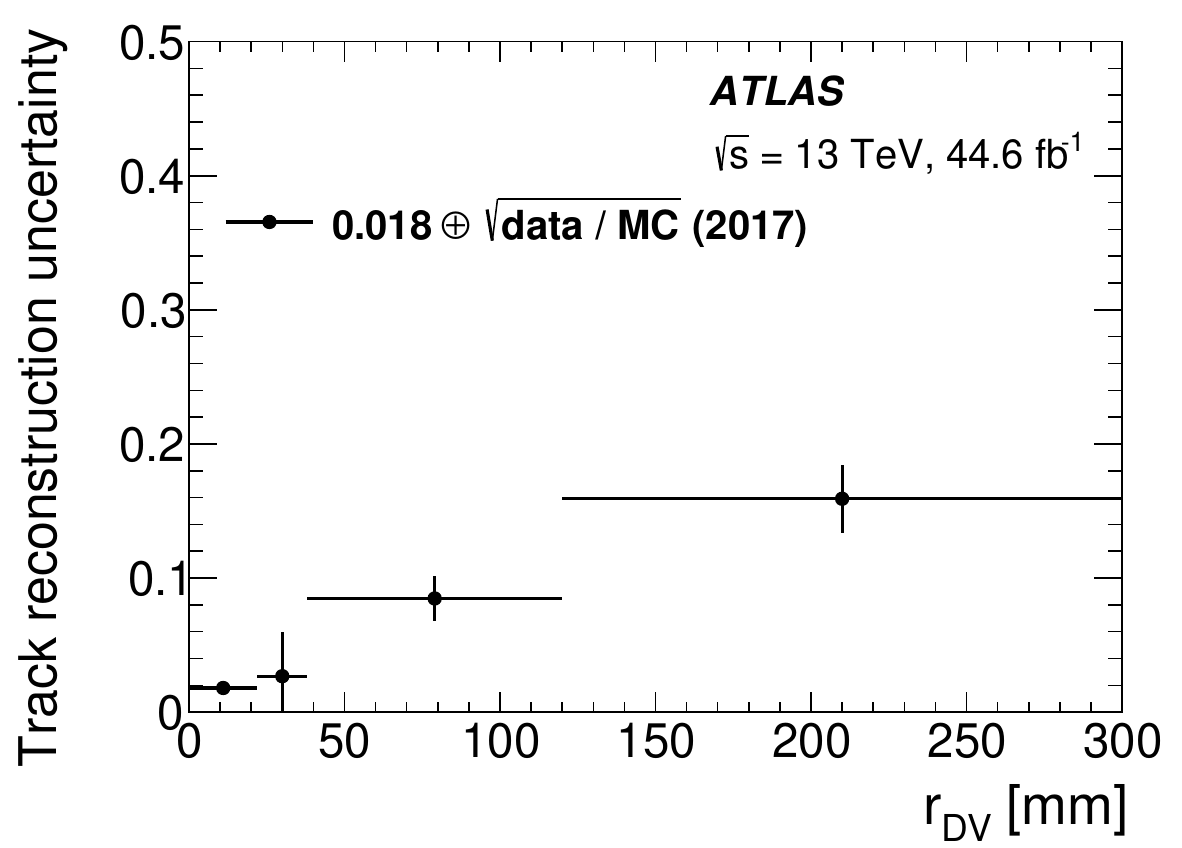}}
\subfloat[\label{fig:tk_2018_sdv_unc}]{\includegraphics[width=0.33\textwidth]{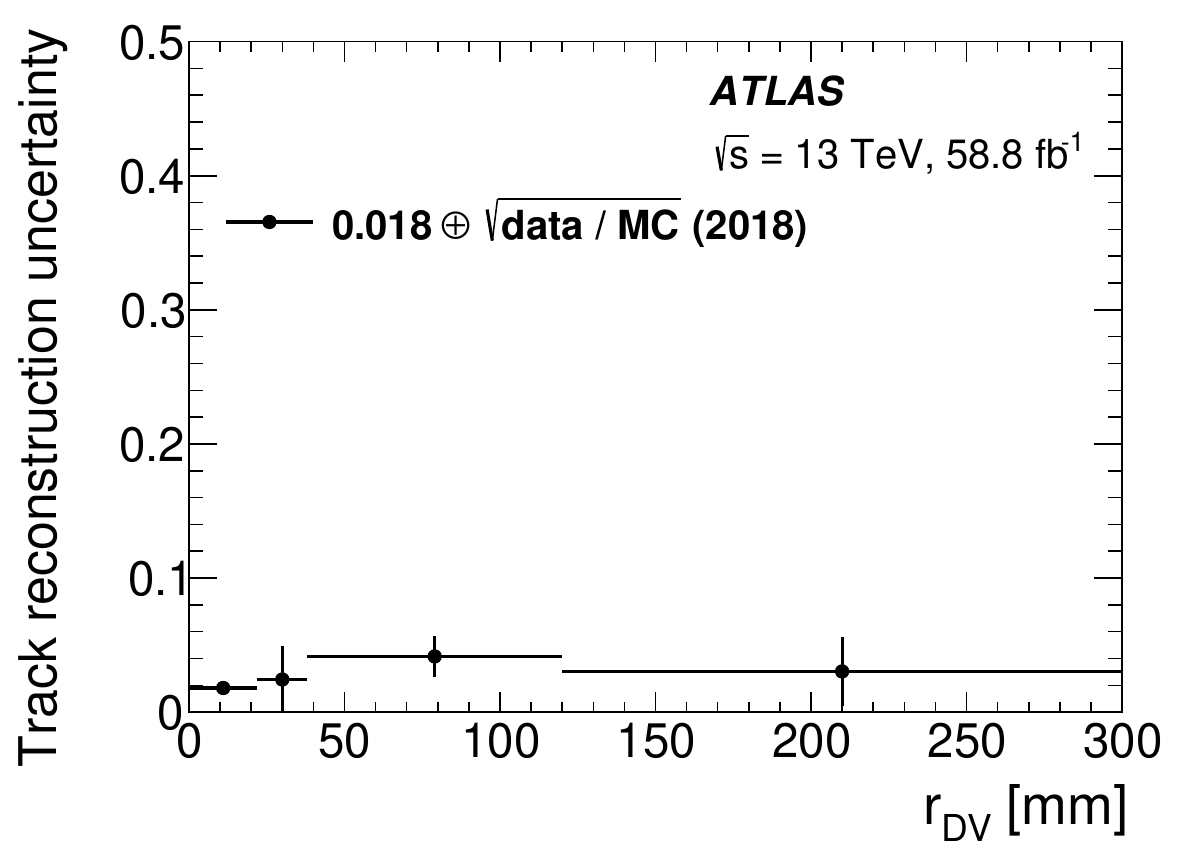}} \\
\caption{
Example counts of $2$-track $\kso \rightarrow \pi^+ \pi^-$ candidate decay SDV in MC and data and their MC/data ratios, for years \protect\subref{fig:tk_2016_sdv_ratio} 2016, \protect\subref{fig:tk_2017_sdv_ratio} 2017 and \protect\subref{fig:tk_2018_sdv_ratio} 2018. The error bars include statistical uncertainty only. The MC counts are normalized to match the data in the innermost bin.
The resulting track reconstruction uncertainties are shown in \protect\subref{fig:tk_2016_sdv_unc}--\protect\subref{fig:tk_2018_sdv_unc}, where a (standard) baseline 1.8\% uncertainty has been added in quadrature.
}
\label{fig:track_killing}
\end{figure}

\begin{figure}[!htb]
\centering
\subfloat[\label{fig:tracking_syst_sdv}]{\includegraphics[width=0.33\textwidth]{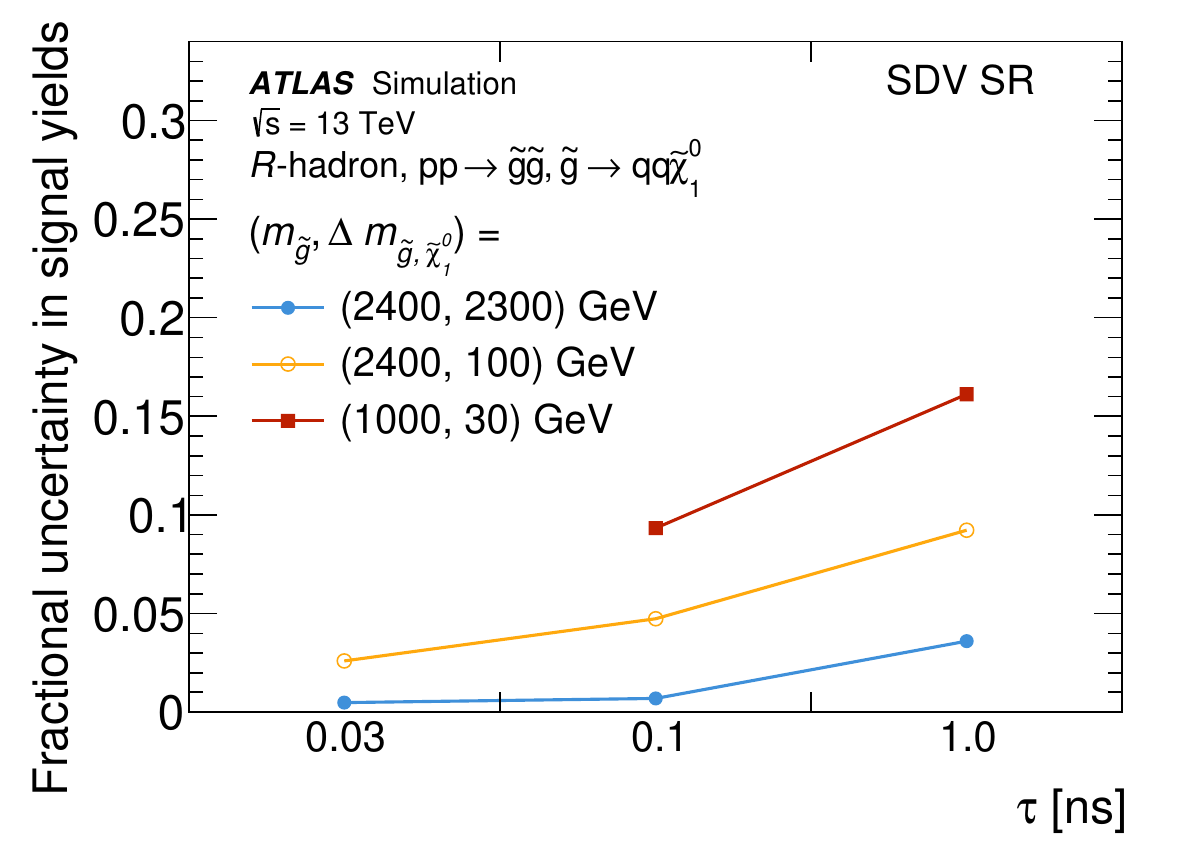}} %
\subfloat[\label{fig:tracking_syst_1fdv}]{\includegraphics[width=0.33\textwidth]{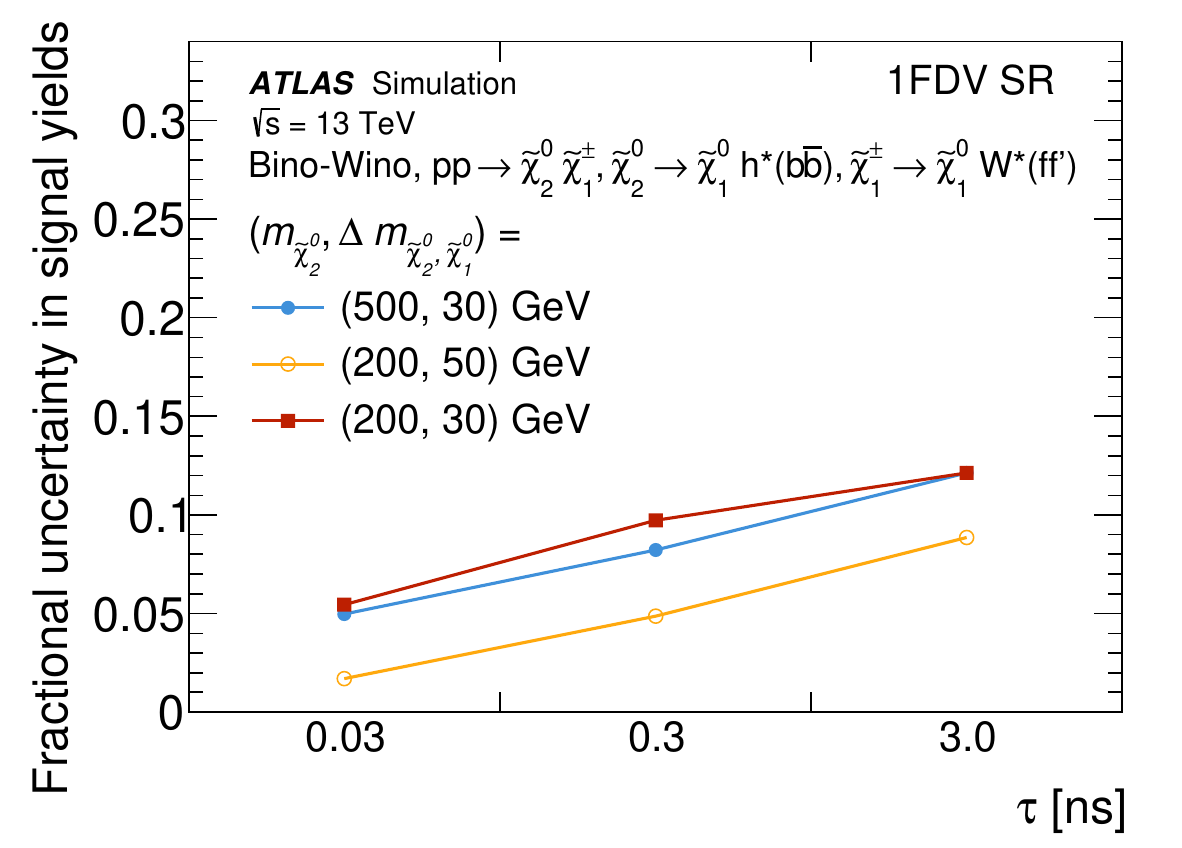}} %
\subfloat[\label{fig:tracking_syst_2fdv}]{\includegraphics[width=0.33\textwidth]{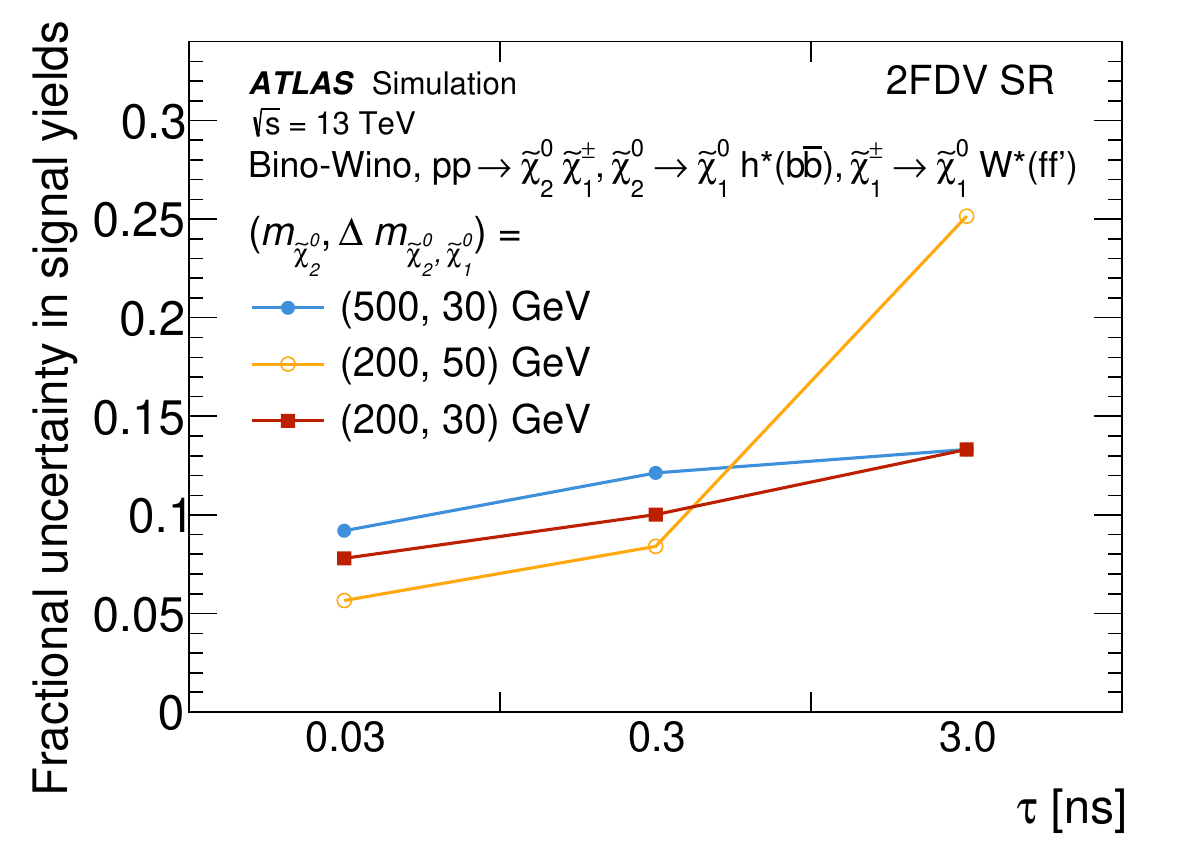}} %
\caption{
Example of fractional uncertainties on the signal yields from track reconstruction uncertainties, for \protect\subref{fig:tracking_syst_sdv} the SDV SR for the \RhadronModel, and \protect\subref{fig:tracking_syst_1fdv} the 1FDV and \protect\subref{fig:tracking_syst_2fdv} the 2FDV SRs for the \BinoWinoModel model.
}
\label{fig:tracking_syst}
\end{figure}

For estimating the uncertainties related to tracking inefficiencies, an MC-data comparison is performed using $2$-track \kso candidate DVs. The \kso candidate DV yield is measured as a function of \rDV. This \rDV-dependent yield profile is then compared between data and MC simulation and interpreted as a proxy for tracking inefficiencies that are imperfectly modeled in the simulation.
These are combined in quadrature with a baseline $1.8\%$ tracking inefficiency, determined from performance studies of standard tracking.
Track reconstruction uncertainties estimated via this method for SDVs are shown in \Fig{\ref{fig:track_killing}}.
The same process is also performed for FDVs.
The obtained FDV associated tracking uncertainties are similar to those of SDV in terms of both the magnitude of the uncertainty and its year or \rDV dependency.
This combined tracking inefficiency estimate is then used to perform a ``track-killing" study, where tracks are removed from DVs (separately for SDVs and FDVs) based on the estimated inefficiencies and the signal yields recomputed.
The track reconstruction uncertainty for some representative signal samples is shown in \Fig{\ref{fig:tracking_syst}}.
The uncertainty in the signal yields from the track reconstruction uncertainty is
one of the dominant sources of systematic uncertainty for most samples, being
particularly pronounced for those corresponding to longer LLP lifetimes, where
DVs increasingly include LRT tracks, which are expected to have larger reconstruction
uncertainty~\cite{ATL-PHYS-PUB-2017-014}. It contributes up to 18\% in the \HiggsPortalModel model.


The signal samples considered are normalized to the recorded luminosity of the 2016-2018 dataset. The uncertainty
in this recorded luminosity, discussed in \Sect{\ref{sec:atlas}}, is propagated as an additional experimental
systematic uncertainty on each signal.

\subsection{Theory uncertainties}
\label{sec:theory_syst}

Theory uncertainties are calculated during the generation of the MC signal samples described in \Sect{\ref{sec:signals}} by varying the choice of renormalization and factorization scales by a factor of two; this approach generates six variations, two in which the scales are varied together and four in which the scales are varied independently.
A seven-point envelope is then taken around the nominal value to evaluate an uncertainty.

The choices of parton distribution function (PDF) set and strong coupling constant $\alpha_s$ are also varied during MC event generation. These uncertainties are then evaluated for each physics signal process and the signal yields in the relevant SRs are then recomputed for each variation. The standard deviation of those yields is taken as the systematic uncertainty for that sample.

Theory uncertainties contribute approximately 0.6--14\% depending on the signal model.
For the \RhadronModel model, the full set of MC event-level theory uncertainties could not
be evaluated for each sample, due to complexities in simulating \rhadrons. A conservative
theory uncertainty is instead applied based on the largest deviation observed amongst a small but representative subset of samples.

\begin{table}[h!]
\centering
\caption{Ranges of relative systematic uncertainties (16th--84th percentile, equivalent to a $\pm 1$ standard deviation if the distributions were Gaussian) for each signal model and associated signal region.\label{tab:systematics}}
\begin{tabular}{l|c|c|c|c}
\toprule
Systematic & \RhadronModel & \AxinoModel, $ZZ$ & \HiggsPortalModel & \BinoWinoModel \\
uncertainty & (SDV SR) (\%) & (SDV SR) (\%) & (2 FDV SR) (\%)& (1 FDV SR) (\%)\\
\midrule
JES and JER & $0.16$--$0.56$ & $0.13$--$0.90$ & $0.94$--$1.9$ & $0.27$--$0.59$ \\
MET soft term & $0.022$--$0.33$ & $0.018$--$0.12$ & $0.16$--$0.39$ & $0.053$--$0.14$ \\
Theory & $12$ & $0.52$--$1.5$ & $5.0$--$13$ & $3.0$--$7.5$ \\
Tracking & $2.1$--$13$ & $0.89$--$7.2$ & $6.4$--$18$ & $2.3$--$9.6$ \\
\bottomrule
\end{tabular}
\end{table}

Table~\ref{tab:systematics} presents the ranges of systematic uncertainties for each signal model, grouped by signal region. The values reflect the spread of uncertainties observed across all tested parameter combinations of each signal model, using the 16th to 84th percentile interval to approximate one standard deviation while retaining asymmetric variation where present.

There are additional theory uncertainties in the production cross-sections, determined by the LHC
SUSY \cite{Beenakker:2014sma, Fuks:2012qx, Fuks:2013vua} and Higgs working groups~\cite{deFlorian:2016spz}.
These uncertainties only impact the normalization of the overall yield, not the selection efficiency,
and so are treated separately and not included in Table~\ref{tab:systematics}.


\section{Results and interpretation}
\label{sec:results}

One event is observed in SDV SR and five events are observed
in FDV SR. Of those five, three pass the 1FDV SR
criteria and two pass the 2FDV SR criteria.
A small local excess is seen over the background prediction in the 1FDV SR
reported in Table \ref{tab:uncertaintySummary},
but in the 2FDV and SDV SRs there is good agreement between data and background.

A series of statistical tests on the observed events are then performed; the statistical model is described in Section \ref{sec:fitting} below,
with results in \Sect{\ref{sec:model_independent}} and \ref{sec:limits}.

%
\subsection{Statistical model}
\label{sec:fitting}

To determine the compatibility of each signal model with the observed data, a
series of profile likelihood fits~\cite{Rolke:2004mj} are performed, with a profile negative-log-likelihood ratio
test statistic~\cite{Cowan:2010js}.
A \CLs frequentist approach~\cite{Read:2002hq} is used to set 95\% confidence-level limits.

Fits are performed by comparing $\mu S + B$ to the observed data in each
signal region. The signal and background yields $S$ and $B$ are modelled
by the product of Poissonian distributions, with the background yield in
the signal region evaluated as described in \Sect{\ref{sec:sr_backgrounds}}.
The signal strength parameter $\mu$ is a free parameter that scales the number
of predicted signal events.
The discovery $p$-value, which gives the consistency of the background-only hypothesis with
the data, can be determined by setting $\mu = 0$.
As the low signal region counts invalidate the asymptotic approximation~\cite{Cowan:2010js}, fits are performed using 10,000 pseudo-experiments to numerically
determine the distribution of the test statistic for a given signal hypothesis.
The software packages \pkg{pyhf} 0.7.6~\cite{pyhf,pyhf_joss_modified} and \pkg{cabinetry} 0.6.0~\cite{cabinetry, cabinetry_paper}
are used to carry out these steps.

The SDV and FDV signal regions are fit independently. A single-bin fit is performed
for the SDV region. The two FDV regions can also be fit independently as single
bins, as is done in \Sect{\ref{sec:model_independent}} below,
or combined into one two-bin fit with the 1FDV and 2FDV sub-regions treated as separate
bins, as is done in \Sect{\ref{sec:limits}}.
Systematic uncertainties on the signal or background are represented by a series
of uncorrelated Gaussian nuisance parameters, although in the combined FDV
fit, each signal systematic is treated as fully correlated across the two FDV
bins. The definitions of all systematics can be found in \Sect{\ref{sec:systematics}} above.
The background estimate systematics are consistently the dominant source of
uncertainty in the fit.

Discovery $p$-values are calculated and quoted in Table \ref{tab:modelIndependentLimits}. Since no significant excess was seen in any of the signal regions, limits
are set on both a generic (model-independent) BSM signal and on the specific
signal models introduced in \Sect{\ref{sec:signals}}. Limits are set by testing different
signal hypotheses, varying the signal strength to find the value of $\mu$
corresponding to a 95\% confidence level. Separate fits were performed
for each signal model in both the SDV and combined FDV regions. Although the optimization
of selection criteria was guided by the signal region efficiency of each model,
as summarized in Table \ref{tab:sigeff_table}, the region
which provides the strictest expected limits on a given model is ultimately used to set
the limits reported in \Sect{\ref{sec:limits}} below.

The theory uncertainties in the production cross-sections for the different signal models are
not treated as nuisance parameters. Instead, fits are performed to the signal yield
scaled up or down by this uncertainty in order to set limits on a $\pm 1 \sigma$
variation of the production cross-section.


%
\subsection{Model-independent limits}
\label{sec:model_independent}

Model-independent constraints are first determined on the presence of excess signal above the estimated backgrounds. The methods of maximal likelihood fitting described above are used to set constraints on the \textit{visible cross-section} of any non-SM process in any of the signal regions, provided such a signal does not contaminate any of the control regions used for background estimation.
To avoid assumptions about how a generic signal would be correlated between the 1FDV and 2FDV regions, separate 1FDV and 2FDV fits are performed. As these fits are performed with a signal, there are no signal systematic uncertainties; the only nuisance parameters included in these fits are for the systematic uncertainties in the background estimates.

The product of the signal strength and $\SI{1}{fb}$ test signal, $\mu S$, is interpreted
as the visible cross-section $\langle A \epsilon \sigma \rangle$,
which is the product of acceptance, detection efficiency and process cross-section.
\Tab{\ref{tab:modelIndependentLimits}} provides the model-independent limits for the SDV and FDV signal regions.
\Sobs and \Sexp are the 95\% CL upper limits on the observed and expected number of signal events under the background-only hypothesis.
The observed limit was found to be within $1 \sigma$ of the expected limit for the SDV signal region,
and within $2 \sigma$ for the FDV signal regions.

\begin{table}[]
\centering
\caption{
Model-independent limits for the SDV and combined FDV signal regions. The observed number
of events, compared to expected background, is also shown for comparison.
\AccEffXsect is the 95 CL upper limit on the visible cross-section,
\Sobs and \Sexp is the 95\% CL upper limits on the observed and expected number of signal events
(the latter given the expected number of background events),
$p(s=0)$ is the discovery $p$-value and
$Z$ is the corresponding Gaussian significance.
}
\begin{tabular}{ c | c c c c c c c c}
\toprule
Region & Observed (Expected) & \AccEffXsect[fb] & \Sobs & \Sexp & $p(s=0)$ & $Z$ \\  %
\midrule
SDV SR & 1 ($0.6 \pm 0.4$) & $0.0280$ & $3.84$ & $3.20^{+1.17}_{-0.05}$  & $0.32$ & $0.46$ \\ %
\midrule
1FDV SR & 3 ($0.8 \pm 0.5$) & $0.0469$ & $6.42$ & $3.62^{+1.53}_{-0.18}$  & $0.06$ & $1.59$ \\ %
2FDV SR  & 2 ($1.3 \pm 0.6$) & $0.0353$ & $4.83$ & $3.98^{+1.59}_{-0.63}$  & $0.31$ & $0.50$ \\ %
\bottomrule
\end{tabular}
\label{tab:modelIndependentLimits}
\end{table}
\subsection{Interpretations}
\label{sec:limits}

The methods described above are also used to set model-dependent limits on each of the four signal models in either the SDV or the combined FDV region.

\begin{figure}[!htb]
\centering
\subfloat[\label{fig:gluino_limits_mg_vs_tau}]{
\includegraphics[width=0.49\textwidth]{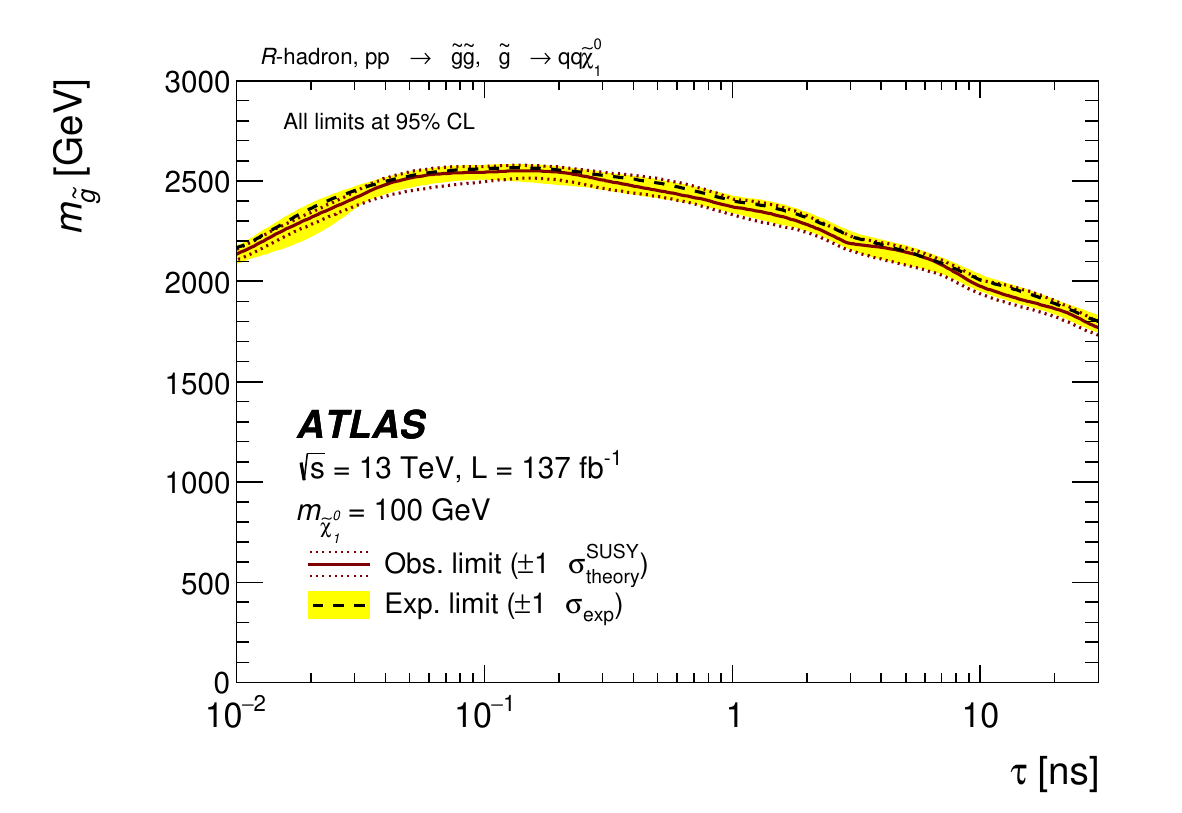}
}
\subfloat[\label{fig:gluino_limits_mchi_vs_mg}]{
\includegraphics[width=0.49\textwidth]{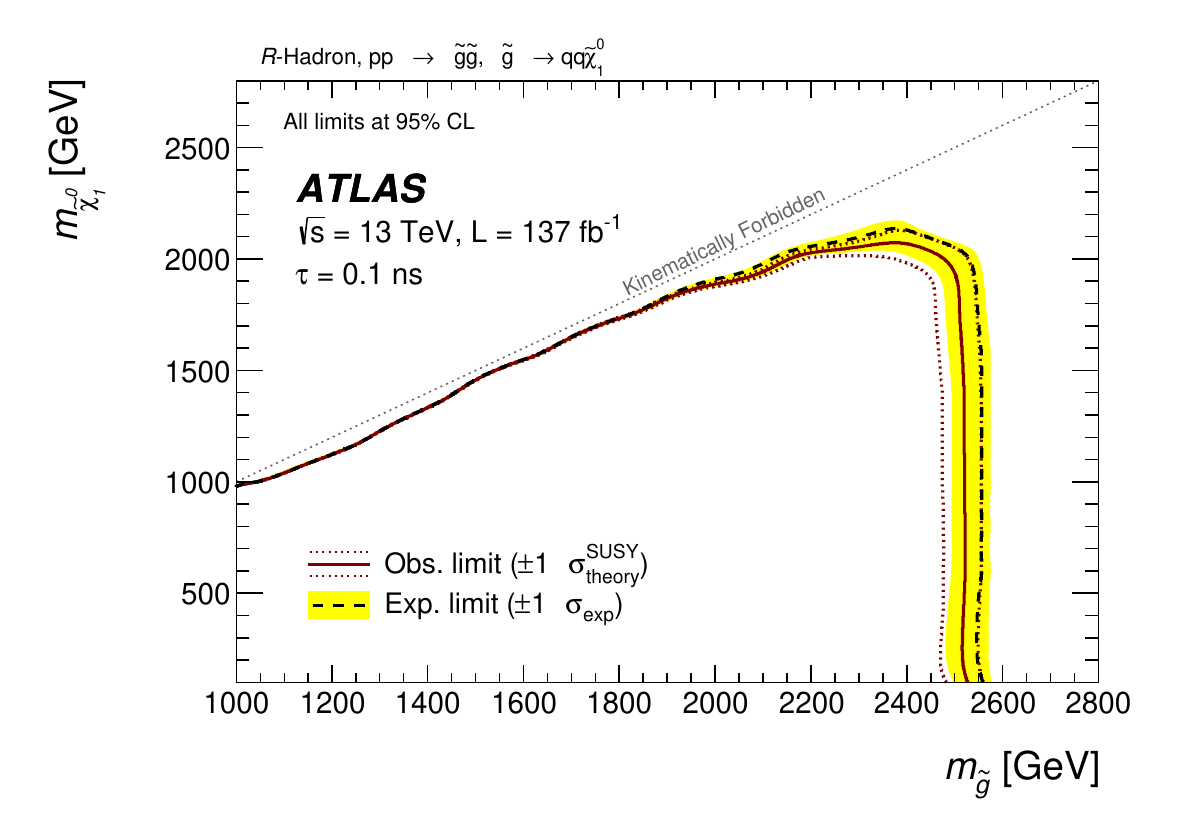}
}
\caption{
Observed and expected 95\% confidence-level upper limits on the \rhadron model as a function of \protect\subref{fig:gluino_limits_mg_vs_tau} the gluino lifetime and \rhadron mass with the lightest neutralino mass $\mneutralino = \SI{100}{GeV}$, and \protect\subref{fig:gluino_limits_mchi_vs_mg} the gluino and neutralino masses for a gluino lifetime $\tau = \SI{0.1}{ns}$.
The area below the contour is excluded.
}
\label{fig:gluino_limits}
\end{figure}

The strongest limits on the \RhadronModel come from the SDV signal region, as expected given the absence of $b$-jets in the final state.
The observed and expected limits are shown in \Fig{\ref{fig:gluino_limits}} as a function of both mass and lifetime, showing that the gluino mass is constrained to be greater than approximately 1.8--2.5~TeV for a wide range of neutralino masses and gluino lifetimes. These results extend constraints across a lifetime range of 0.01--0.1~ns by more than 200~GeV in \mgluino compared to the 2016-only result from ATLAS~\cite{SUSY-2016-08}.

\begin{figure}[!htb]
\centering
\subfloat[\label{fig:binowino_limits_mchi_vs_tau}]{
\includegraphics[width=0.49\textwidth]{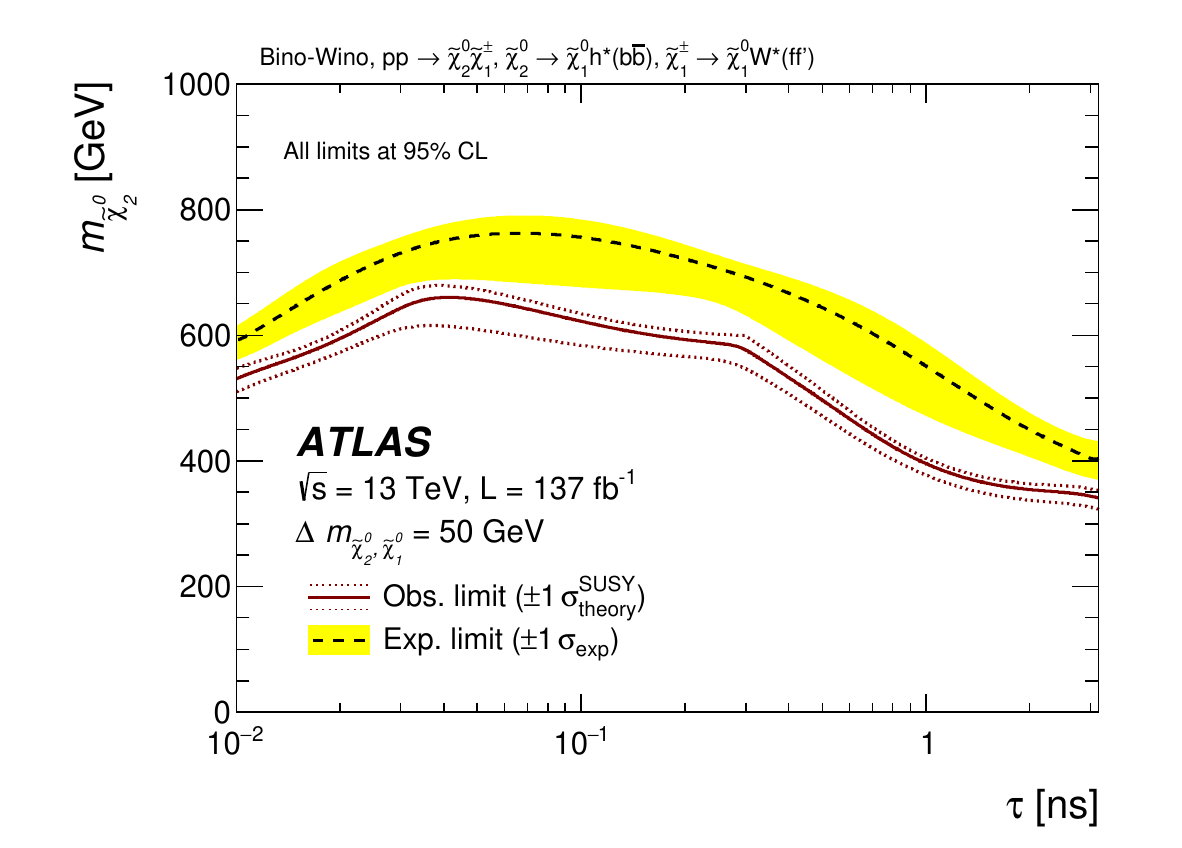}
}
\subfloat[\label{fig:binowino_limits_dmchi_vs_mchi}]{
\includegraphics[width=0.49\textwidth]{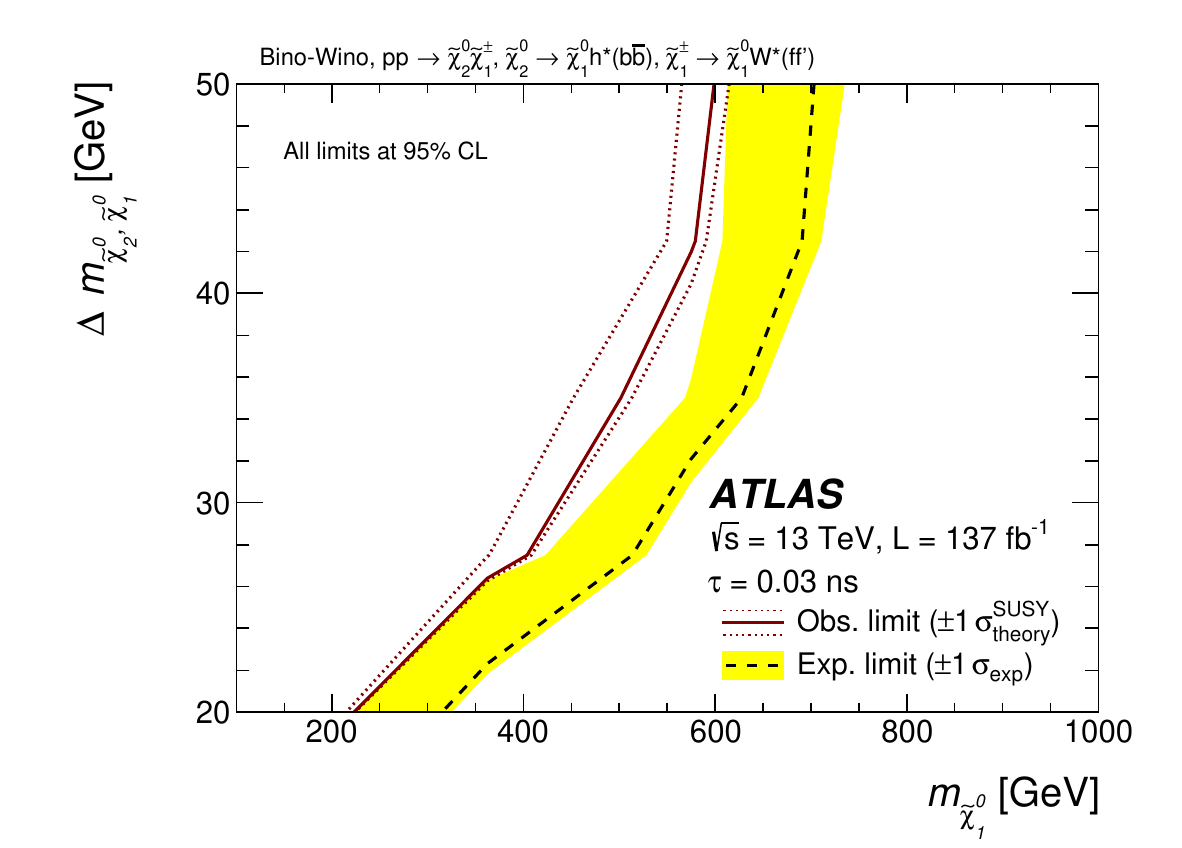}
}
\caption{95\% confidence level observed and expected upper limits on the masses and lifetimes
of the \BinoWinoModel model, with exclusion contours as described in \Fig{\ref{fig:gluino_limits}}.
The exclusion as a function \protect\subref{fig:binowino_limits_mchi_vs_tau} of \ninotwo lifetime and mass with a mass splitting between
\ninotwo and \ninoone $\dMassWinoBino = \gev{50}$ and \protect\subref{fig:binowino_limits_dmchi_vs_mchi} of
mass splitting and \ninoone mass with a lifetime $\tau = \ns{0.03}$. The weaker constraint on the observed limit compared to the expected limit is due to a small excess in FDV SR.
The area below the contour is excluded for the left plot, and the area to the left of the contour is excluded for the right plot.
}
\label{fig:binowino_limits}
\end{figure}

The \BinoWinoModel model described in \Sect{\ref{sec:bino_wino}} performs best in the FDV signal region,
as expected due to the $b$-jets in the final state.
\Fig{\ref{fig:binowino_limits}} shows the observed and expected limits, which constrain the lightest neutralino masses to be greater than approximately 350--650~GeV for a wide range of lifetimes and mass splittings.
The observed limit is lower than the expected limit because of the small excess in 1FDV SR.
The largest observed $p$-value under the \BinoWinoModel model is obtained as 0.52 when $\mslneutralino = \gev{700}, \dMassWinoBino = \gev{40}$, and $\tau$ = 0.3~ns.

\begin{figure}[!htb]
\centering
\subfloat[\label{fig:axino_limits_xsect_vs_maxion}]{
\includegraphics[width=0.485\textwidth]{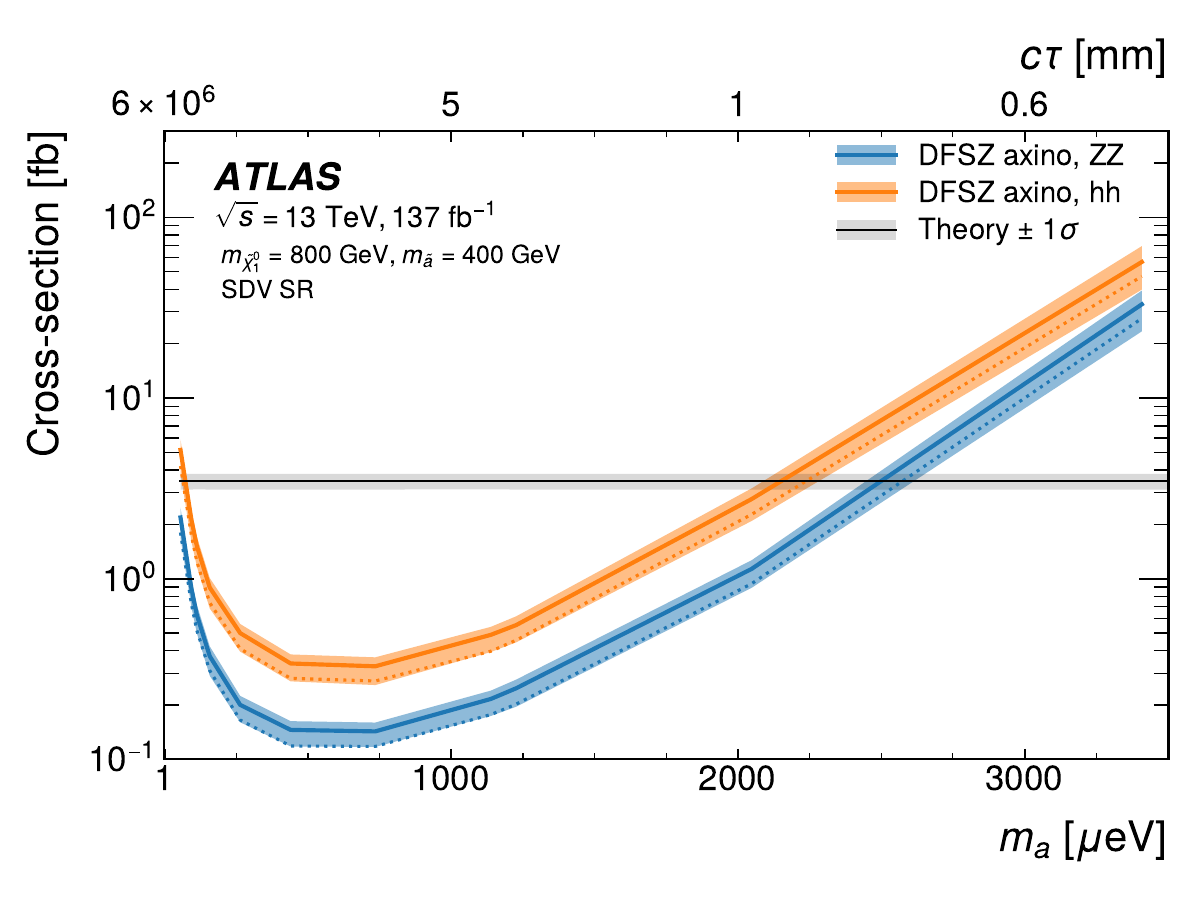}
}
\subfloat[\label{fig:axino_limits_xsect_vs_maxino}]{
\includegraphics[width=0.50\textwidth]{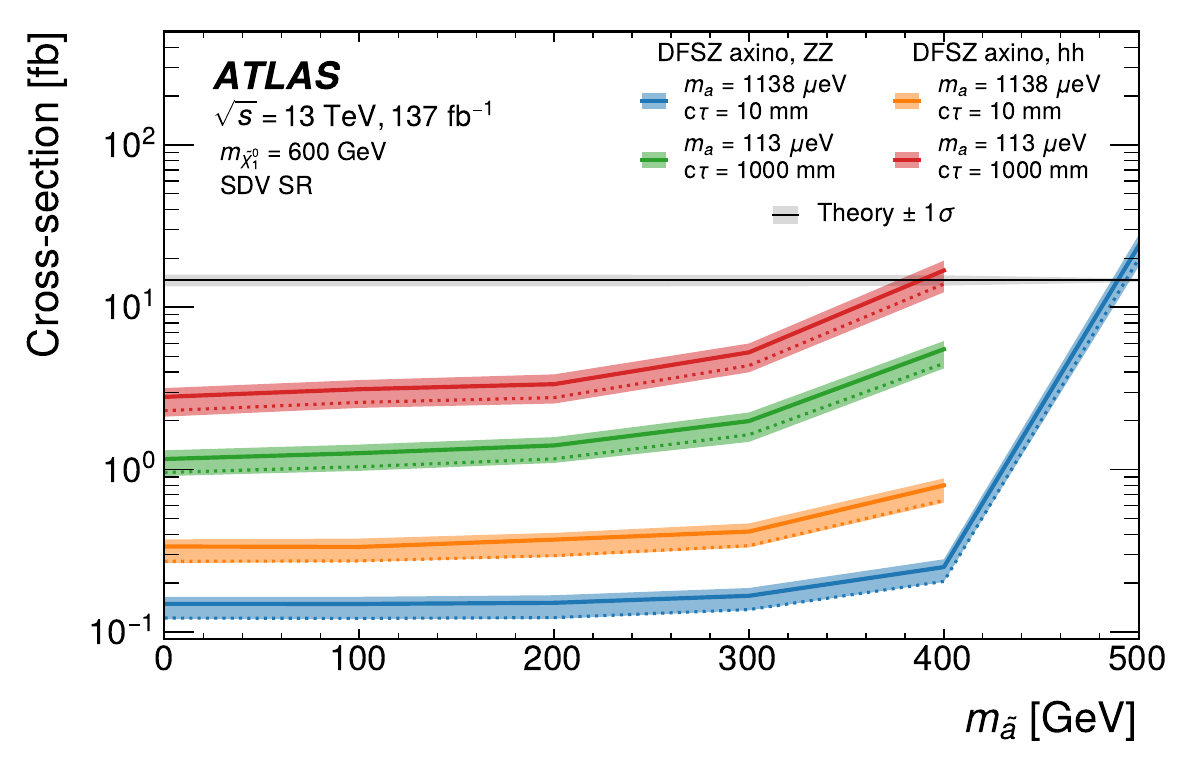}
}
\caption{Expected and observed limits at 95\% confidence level on the cross-section of the \AxinoModel model as a function of axion and axino mass, in which either \protect\subref{fig:axino_limits_xsect_vs_maxion} the axino mass \maxino is held constant or \protect\subref{fig:axino_limits_xsect_vs_maxino} the neutralino
mass \mneutralino is held constant. Limits are drawn separately for two choices of the axion mass parameter \maxion, which correspond to a range
of \ninoone lifetimes as discussed in \Sect{\ref{sec:axino}}. The theoretical Higgsino pair production cross-section used is included for reference \cite{Fuks:2012qx, Fuks:2013vua}.
Cross-sections above the contours are excluded.
}
\label{fig:axino_limits}
\end{figure}

In the \AxinoModel model described in \Sect{\ref{sec:axino}}, the final state can include either
four displaced $b$-jets (in the case of a Higgs boson final state) or four displaced light-quark jets (in the case of a $Z$ boson final state). In principle, the presence of $b$-jets in the final state should lead to stronger sensitivity in the FDV SR for the Higgs boson scenario, while the SDV SR should perform better for the $Z$ boson final state. However, stronger expected limits were seen for both final states in the SDV SR than in the FDV SR\footnote{
The FDV SR is optimized for the \BinoWinoModel case, and differs in kinematics and range of parameters
DFSZ axino Higgs channel in several ways. For all axino
models studied, the Higgs boson is required to be produced on-shell, while for the
\BinoWinoModel models the Higgs boson is instead produced off-shell. This explains why this SR
does not display an optimal sensitivity for the axino Higgs channel.
}, with the $Z$ boson final state providing more stringent constraints as expected. As shown in \Fig{\ref{fig:axino_limits}}, the observed cross-section limits for the $Z$ boson final state are 2--3 times better than those for the Higgs boson final state, consistent with the better performance of
the SDV algorithm in reconstructing the light-quark jets from $Z$ decays. The constraints exclude significant portions of the studied parameter space at 95\% confidence level for both production mechanisms, particularly for axion masses (\maxion) between 100 and 2500~$\mu$eV when the neutralino mass is fixed at 800~GeV \protect\subref{fig:axino_limits_xsect_vs_maxion}, and for neutralino masses (\mneutralino) below approximately 400~GeV when
the axion mass parameters are held constant \protect\subref{fig:axino_limits_xsect_vs_maxino}. This is the first time limits have been set explicitly on this model; while a past search
has probed a similar DFSZ axino scenario~\cite{SUSY-2018-41}, it was previously assumed that the axino would be promptly produced and did not directly constrain \fa.

\begin{figure}[!htb]
\centering
\includegraphics[width=0.60\textwidth]{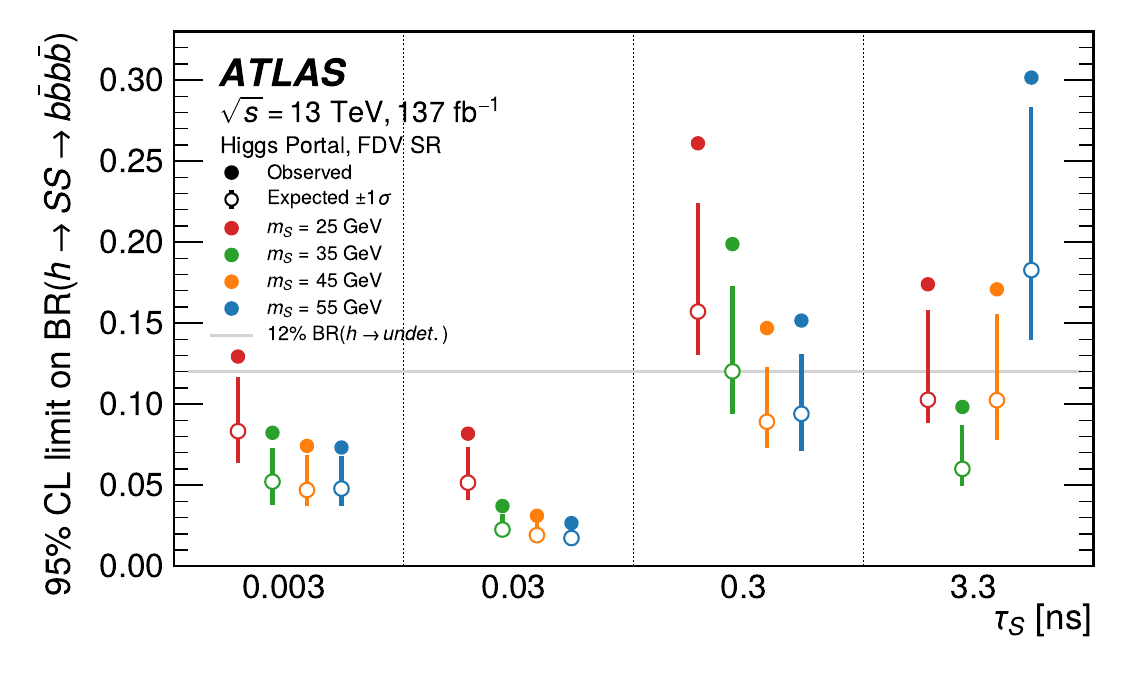}
\caption{Expected and observed limits on the branching ratio of $h \to SS \to 4b$ in the Higgs
portal model at 95\% confidence level, shown as a function of the lifetime $\tau$ of the new
heavy \scalar $S$. Limits are evaluated for several values of the \scalar mass \mS,
and compared to a 12\% branching ratio for the Higgs boson decaying into BSM
final states \cite{HIGG-2021-23}. Each color indicates a specific value of \mS.
The weaker constraint on the observed limit compared to the expected limit is due to a small excess in FDV SR.}
\label{fig:hp_limits}
\end{figure}

The Higgs portal model introduced in Section~\ref{sec:higgs_portal} is best constrained in the FDV SR. Observed and expected limits are set on the branching ratio of $h \to SS \to 4b$ and shown in \Fig{\ref{fig:hp_limits}}. These exclusions are compared to a target 12\% branching ratio, above which exotic undetected Higgs boson decays have been excluded \cite{HIGG-2021-23}. For \scalar masses $\mS = \gev{25}$ and above, this analysis excludes a significant
region of phase space below this target, down to a branching ratio of 2.6--12.9\% for \scalar lifetimes $\tau_{S} \le \ns{0.03}$. Complementary searches for LLPs place stronger constraints on the Higgs portal model for lifetimes of 0.03 and 0.3 ns \cite{EXOT-2021-32,CMS-EXO-21-008,EXOT-2022-04,EXOT-2022-17,CMS-EXO-23-013,ATL-PHYS-PUB-2025-039}. %

The shape of the exclusion contours in Figures~\ref{fig:gluino_limits}–\ref{fig:hp_limits} reflects the interplay between displaced vertex reconstruction efficiency and LLP decay kinematics. Sensitivity peaks at intermediate lifetimes, where decays are typically well-contained within the tracking volume for vertices with a sufficiently high \mDV that pass all selection and cleaning requirements. At low \mDV, signal vertices are more likely to mimic common background topologies and are frequently removed by the cleaning cuts. At large lifetimes, LLPs increasingly decay outside the detector volume, reducing the probability of reconstructing a displaced vertex. In models predicting two displaced decays per event, some sensitivity is retained even at long lifetimes when one LLP decays within the tracker and the other escapes, contributing to the event's \met. However, such events still require at least one well-reconstructed displaced vertex to satisfy the analysis selection, and the overall efficiency drops sharply in this regime.



%
%
%
\FloatBarrier

\section{Conclusion}
\label{sec:conclusion}

The results of a multi-faceted search for BSM physics processes characterized by displaced vertices and missing transverse momentum using \invfb{137} of proton--proton collision data delivered by the LHC and collected by the ATLAS detector at $\sqs = \tev{13}$ are presented. This analysis employed both inclusive event selection criteria and dedicated displaced vertex reconstruction algorithms to maximize sensitivity across a broad range of theoretical models predicting long-lived particle signatures.

No statistically significant excess above the SM background expectation was observed, thus allowing constraints to be placed on the four primary BSM signal models considered: long-lived gluinos which hadronize into \rhadrons~before decaying, neutralinos decaying via Higgs-mediated channels, long-lived Higgsinos decaying to axinos, and
an exotic Higgs portal model predicting displaced decays of light \scalars.

In the \RhadronModel model, constraints reached gluino masses of 1.8--2.5~TeV for a range of neutralino masses and gluino lifetimes, improving previous ATLAS gluino mass results by more than 200~GeV.
In the \BinoWinoModel model, the lightest neutralino masses are constrained to be greater than approximately 350--650~GeV for a range of lifetimes and mass splittings.
In the \AxinoModel model the analysis constrains axion masses between 100 and 2500~$\mu$eV when the neutralino mass is fixed at 800~GeV, and for neutralino masses below 400~GeV for fixed axion mass parameters. Finally, the excluded Higgs branching ratio in the \HiggsPortalModel model reaches 2.6\% for a \scalar with a mass of 55~GeV and lifetime of \ns{0.03}.

%


%
%
%
%
%
\section*{Acknowledgements}
%

%

%
%

%
%

We thank CERN for the very successful operation of the LHC and its injectors, as well as the support staff at
CERN and at our institutions worldwide without whom ATLAS could not be operated efficiently.

The crucial computing support from all WLCG partners is acknowledged gratefully, in particular from CERN, the ATLAS Tier-1 facilities at TRIUMF/SFU (Canada), NDGF (Denmark, Norway, Sweden), CC-IN2P3 (France), KIT/GridKA (Germany), INFN-CNAF (Italy), NL-T1 (Netherlands), PIC (Spain), RAL (UK) and BNL (USA), the Tier-2 facilities worldwide and large non-WLCG resource providers. Major contributors of computing resources are listed in Ref.~\cite{ATL-SOFT-PUB-2026-001}.

We gratefully acknowledge the support of ANPCyT, Argentina; YerPhI, Armenia; ARC, Australia; BMWFW and FWF, Austria; ANAS, Azerbaijan; CNPq and FAPESP, Brazil; NSERC, NRC and CFI, Canada; CERN; ANID, Chile; CAS, MOST, NSFC and FRFCU, China; Minciencias, Colombia; MEYS CR, Czech Republic; DNRF and DNSRC, Denmark; IN2P3-CNRS and CEA-DRF/IRFU, France; SRNSFG, Georgia; BMFTR, HGF and MPG, Germany; GSRI, Greece; RGC and Hong Kong SAR, China; ICHEP and Academy of Sciences and Humanities, Israel; INFN, Italy; MEXT and JSPS, Japan; CNRST, Morocco; NWO, Netherlands; RCN, Norway; MNiSW, Poland; FCT, Portugal; MNE/IFA, Romania; MSTDI, Serbia; MSSR, Slovakia; ARIS and MVZI, Slovenia; DSI/NRF, South Africa; MICIU/AEI, Spain; SRC and Wallenberg Foundation, Sweden; SERI, SNSF and Cantons of Bern and Geneva, Switzerland; NSTC, Taipei; TENMAK, T\"urkiye; STFC/UKRI, United Kingdom; DOE and NSF, United States of America.

Individual groups and members have received support from BCKDF, CANARIE, CRC and DRAC, Canada; CERN-CZ, FORTE and PRIMUS, Czech Republic; COST, ERC, ERDF, Horizon 2020 and Marie Sk{\l}odowska-Curie Actions, European Union; Investissements d'Avenir Labex, Investissements d'Avenir Idex and ANR, France; DFG and AvH Foundation, Germany; Herakleitos, Thales and Aristeia programmes co-financed by EU-ESF and the Greek NSRF, Greece; BSF-NSF and MINERVA, Israel; NCN and NAWA, Poland; La Caixa Banking Foundation, CERCA and AGAUR programs from Generalitat de Catalunya and PROMETEO and GenT Programmes Generalitat Valenciana, Spain; G\"{o}ran Gustafssons Stiftelse, Sweden; The Royal Society and Leverhulme Trust, United Kingdom; Eric and Wendy Schmidt Fund for Strategic Innovation, United States of America.

In addition, individual members wish to acknowledge support from Chile: Agencia Nacional de Investigaci\'on y Desarrollo (ANID FONDECYT reg. 1230987, FONDECYT 1230812, FONDECYT 1240864, Fondecyt 3240661, Fondecyt Regular 1240721); China: Fundamental Research Funds for the Central Universities (010-63263105), Chinese Ministry of Science and Technology (MOST-2023YFA1605700, MOST-2023YFA1609300), National Natural Science Foundation of China (NSFC 12275265, NSFC-W2543005); Czech Republic: Czech Science Foundation (GACR - 24-11373S), Ministry of Education Youth and Sports (ERC-CZ-LL2327, FORTE CZ.02.01.01/00/22\_008/0004632), PRIMUS Research Programme (PRIMUS/21/SCI/017); EU: H2020 European Research Council (ERC - 101002463); European Union: European Research Council (BARD No. 101116429, ERC - 101219398, ERC - 948254, ERC 101089007), European Regional Development Fund (HE COFUND GA No.101081355, ERDF), Marie Sklodowska-Curie Actions (GAP-101168829); France: Agence Nationale de la Recherche (ANR-21-CE31-0013, ANR-22-EDIR-0002, ANR-24-CE31-0504-01); Germany: Deutsche Forschungsgemeinschaft (DFG - 469666862); China: Research Grants Council (GRF); Italy: Ministero dell'Università e della Ricerca (NextGenEU 153D23001490006 M4C2.1.1, NextGenEU I53D23000820006 M4C2.1.1, NextGenEU I53D23001490006 M4C2.1.1, SOE2024\_0000023); Japan: Japan Society for the Promotion of Science (JSPS KAKENHI  JP25H0063, JSPS KAKENHI JP22H01227, JSPS KAKENHI JP22H04944, JSPS KAKENHI JP22KK0227, JSPS KAKENHI JP24K23939, JSPS KAKENHI JP24KK0251, JSPS KAKENHI JP25H00650, JSPS KAKENHI JP25H01291, JSPS KAKENHI JP25K01011, JSPS KAKENHI JP25K01023); Norway: Research Council of Norway (RCN-314472); Poland: Polish National Science Centre (NCN 2021/42/E/ST2/00350, NCN OPUS 2023/51/B/ST2/02507, NCN OPUS nr 2022/47/B/ST2/03059, NCN UMO-2019/34/E/ST2/00393, UMO-2022/47/O/ST2/00148, UMO-2023/49/B/ST2/04085, UMO-2023/51/B/ST2/00920, UMO-2024/53/N/ST2/00869); Spain: Agència de Gestió d'Ajuts Universitaris i de Recerca. (AGAUR - 2023 BP 00141), Ministry of Science and Innovation (RYC2019-028510-I, RYC2020-030254-I, RYC2021-031273-I, RYC2022-038164-I), Ministerio de Ciencia, Innovación y Universidades/Agencia Estatal de Investigaci\'on (EU NextGenerationEU (PRTR-C17.I1), PID2022-142604OB-C22); Sweden: Carl Trygger Foundation (Carl Trygger Foundation CTS 22:2312), Swedish Research Council (Swedish Research Council 2023-04654, VR 2021-03651, VR 2022-03845, VR 2022-04683, VR 2023-03403, VR 2024-05451, VR 2025-05940), Knut and Alice Wallenberg Foundation (KAW 2023.0366); Switzerland: Swiss National Science Foundation (SNSF - PCEFP2\_194658); United Kingdom: The Binks Trust, Royal Society (NIF-R1-231091); United States of America: U.S. Department of Energy (ECA DE-AC02-76SF00515), John Templeton Foundation (John Templeton Foundation 63206), Neubauer Family Foundation.

%
%


%
%
%
%
%
%
%
%
%
%
\FloatBarrier

\printbibliography
\clearpage
 
\begin{flushleft}
\hypersetup{urlcolor=black}
{\Large The ATLAS Collaboration}

\bigskip

\AtlasOrcid[0000-0002-6665-4934]{G.~Aad}$^\textrm{\scriptsize 102}$,
\AtlasOrcid[0000-0001-7616-1554]{E.~Aakvaag}$^\textrm{\scriptsize 17}$,
\AtlasOrcid[0000-0002-5888-2734]{B.~Abbott}$^\textrm{\scriptsize 121}$,
\AtlasOrcid[0000-0002-0287-5869]{S.~Abdelhameed}$^\textrm{\scriptsize 83b}$,
\AtlasOrcid[0000-0002-1002-1652]{K.~Abeling}$^\textrm{\scriptsize 54}$,
\AtlasOrcid[0000-0001-5763-2760]{N.J.~Abicht}$^\textrm{\scriptsize 48}$,
\AtlasOrcid[0000-0002-8496-9294]{S.H.~Abidi}$^\textrm{\scriptsize 30}$,
\AtlasOrcid[0009-0003-6578-220X]{M.~Aboelela}$^\textrm{\scriptsize 44}$,
\AtlasOrcid[0000-0002-9987-2292]{A.~Aboulhorma}$^\textrm{\scriptsize 36e}$,
\AtlasOrcid[0000-0001-5329-6640]{H.~Abramowicz}$^\textrm{\scriptsize 154}$
% ...
\bigskip
\\

% ...
$^{17}$Department for Physics and Technology, University of Bergen, Bergen; Norway.\\
% ...
$^{30}$Physics Department, Brookhaven National Laboratory, Upton NY; United States of America.\\
% ...
$^{36}$$^{(a)}$Facult\'e des Sciences Ain Chock, Universit\'e Hassan II de Casablanca;$^{(b)}$Facult\'{e} des Sciences, Universit\'{e} Ibn-Tofail, K\'{e}nitra;$^{(c)}$Facult\'e des Sciences Semlalia, Universit\'e Cadi Ayyad, LPHEA-Marrakech;$^{(d)}$LPMR, Facult\'e des Sciences, Universit\'e Mohamed Premier, Oujda;$^{(e)}$Facult\'e des sciences, Universit\'e Mohammed V, Rabat;$^{(f)}$Institute of Applied Physics, Mohammed VI Polytechnic University, Ben Guerir; Morocco.\\
% ...
$^{44}$Physics Department, Southern Methodist University, Dallas TX; United States of America.\\
% ...
$^{48}$Fakult\"{a}t Physik, Technische Universit{\"a}t Dortmund, Dortmund; Germany.\\
% ...
$^{54}$II. Physikalisches Institut, Georg-August-Universit\"{a}t G\"ottingen, G\"ottingen; Germany.\\
% ...
$^{83}$$^{(a)}$Khalifa University of Science and Technology, Abu Dhabi;$^{(b)}$New York University Abu Dhabi, Abu Dhabi;$^{(c)}$United Arab Emirates University, Al Ain;$^{(d)}$University of Sharjah, Sharjah; United Arab Emirates.\\
% ...
$^{102}$CPPM, Aix-Marseille Universit\'e, CNRS/IN2P3, Marseille; France.\\
% ...
$^{121}$Homer L. Dodge Department of Physics and Astronomy, University of Oklahoma, Norman OK; United States of America.\\
% ...
$^{154}$Raymond and Beverly Sackler School of Physics and Astronomy, Tel Aviv University, Tel Aviv; Israel.\\
% ...
$^{a}$ Also at Affiliated with an institute formerly covered by a cooperation agreement with CERN.\\
$^{b}$ Also at An-Najah National University, Nablus; Palestine.\\
$^{c}$ Also at Borough of Manhattan Community College, City University of New York, New York NY; United States of America.\\
$^{d}$ Also at Center for Interdisciplinary Research and Innovation (CIRI-AUTH), Thessaloniki; Greece.\\
$^{e}$ Also at Centre of Physics of the Universities of Minho and Porto (CF-UM-UP); Portugal.\\
$^{f}$ Also at CERN, Geneva; Switzerland.\\
% ...

\end{flushleft}


%

%
%

%

%
%
%
%

\end{document}